\documentclass[11pt]{article}

\usepackage{amssymb}
\usepackage{epsfig}
\usepackage{graphics}
\usepackage{graphicx}
\usepackage{color}

\usepackage[latin1]{inputenc}
\usepackage[english]{babel}

\usepackage{amssymb}
\usepackage{amsmath}
\usepackage{amsfonts}

\numberwithin{equation}{section}

\input amssym.def
\input amssym.tex

\usepackage{booktabs}
\usepackage{ctable}
\usepackage{subfigure}
\usepackage{float}
\usepackage{layout}
\usepackage{fancyhdr}
\usepackage{color}
\usepackage{eurosym}
\usepackage{wrapfig}  

\usepackage{xr}

\def\I{\mathrm{1\!l}}

\def\R{{\mathbb R}}
\def\E{{\mathbb E}}

\def\T{{\cal{T}}}
\def\<{\langle}
\def\>{\rangle}

\def\S{{\mathcal S}}

\newtheorem{theorem}{Theorem}[section]

\newtheorem{remark}[theorem]{Remark}

\newcommand{\Dy}{\Delta y}

\def \Frac{\displaystyle\frac}

\begin{document}

\title{A hybrid tree/finite-difference approach for\\Heston-Hull-White type models}

\author{{\sc Maya Briani}\thanks{%
Istituto per le Applicazioni del Calcolo, CNR Roma - {\tt m.briani@iac.cnr.it}}\\
{\sc Lucia Caramellino}\thanks{%
Dipartimento di Matematica,
Universit\`a di Roma Tor Vergata - {\tt caramell@mat.uniromP2.it}}\\
{\sc Antonino Zanette}\thanks{%
Dipartimento di Scienze
Economiche e Statistiche,
Universit\`a di Udine - {\tt antonino.zanette@uniud.it}}}


\maketitle

\begin{abstract}\noindent{\parindent0pt
We study a hybrid tree/finite-difference method
which permits to obtain efficient and accurate
European and American option prices in the Heston Hull-White and
Heston Hull-White2d models. Moreover, as a by-product, we provide a new
simulation scheme to be used for Monte Carlo evaluations. Numerical results show the reliability and the efficiency of the proposed methods.}
\end{abstract}




\section{Introduction}
In this paper we consider the Heston-Hull-White model, which is a joint evolution for the equity value with a Heston-like stochastic volatility and a generalized Hull-White stochastic
interest rate model which is consistent with the term structure of
the interest rates. We consider a further situation where the dividend rate is
stochastic, a case which is called here the ``Heston Hull-White2d model'' and can be of interest in the multi-currency (the dividend rate being interpreted as a further interest rate).
We concern the problem of numerically pricing European and American options written on these models.

At the present time, the literature on this subject is quite poor and includes Fourier-Cosine
methods, semi-closed approximations and finite-difference methods to price vanilla options.
In \cite{GO},  Grzelak and Oosterlee introduce two approximations of the
non-affine models.
The Fourier-Cosine method is then used on this approximating affine
model.
The authors remark that for accurate modeling of hybrid derivatives
it is necessary to be able to describe a non-zero correlation between the processes driving the equity and the interest rate. This is possible in the approximations
presented in their paper but only using approximated affine models.
Haentjens and  in't Hout propose in \cite{hh}  a finite-difference
Alternating Direction Implicit (ADI) scheme for pricing European
options solving the original  three-dimensional Heston-Hull-White
Partial Differential Equation (hereafter PDE).
The Heston Hull-White2d model is treated using semi-closed
approximations in the Foreign Exchange model \cite{go12}.

In this paper, we generalize the hybrid tree/finite-difference approach that has been introduced for the Heston model in the paper \cite{bcz}. In practice, this means to write down an algorithm to price European and American options by means of a backward induction that works following a finite-difference PDE method in the direction of the share process and following a recombining binomial tree method in the direction of the other random sources (volatility, interest rate and possibly dividend rate).

It is well known that there is a link between tree methods and finite-difference methods.
The most remarkable benefits in using recombining binomial trees (let us stress
the terms ``recombining'' and ``binomial'': just two possible recombining jumps at each time-step for each component) are the simplicity of the implementation, the low
computational costs and the efficiency of the output numerical
results. In dimension 1,  one can always build a recombining binomial tree (see
e.g. Nelson and Ramaswamy \cite{nr}) but this is not the case in
multidimensional problems. For example, in the  standard (dimension 2)
Heston model it is not possible to write down a recombining binomial approximating
tree - roughly speaking, this follows from the fact that it is not
possible to produce a function of the Heston components such that the
associated Stochastic Differential Equation (SDE) has a diagonal diffusion coefficient.
A binomial tree approximation for the standard Heston model has been proposed by
Vellekoop and Nieuwenhuis in \cite{vn} but it is far from being  recombining and, as shown in   \cite{bcz}, it is problematic when the Feller condition is not satisfied. Finally, an approximation of the price coming from a numerical treatment of the (multidimensional) PDE can be very expensive, mainly to handle the 4-dimensional Heston-Hull-White2d model.

So, the idea underlying the approach developed in this paper is in some sense very simple: we apply the most efficient and easy to implement method whenever we can do it. In fact, wherever an efficient recombining binomial tree scheme can be settled (volatility, interest rate and possibly dividend rate), we use it. And where it cannot (share process), we use a standard (and efficient, being in dimension 1) numerical PDE approach. Hence we avoid to work with expensive (because non  recombining and/or binomial) trees or with PDEs in high dimension. Moreover, for the Cox-Ingersoll-Ross (hereafter CIR) volatility component, we apply the recombining binomial tree method firstly introduced in \cite{bib:acz}, which theoretically converges and efficiently works in practice also when the Feller condition fails.

The description of the approximating processes coming from our hybrid tree/finite-difference approach, suggests a simple way to simulate paths from the Heston-Hull-White models. Therefore, we propose here also a new Monte Carlo algorithm for pricing options which seems to be a real alternative to the Monte Carlo method that makes use of the efficient simulations provided by Alfonsi \cite{al}.

Our approaches allow one to price options in the
original Heston-Hull-White processes with non-zero correlations. Here, we consider the case of a non null correlation  between
the equity and the interest rate process, as well as between the equity and the stochastic volatility. Moreover, in the Heston-Hull-White2d model, we allow the dividend rate to be stochastic and correlated to the equity process. But it is worth noticing that other sets of correlations can surely be selected.

\smallskip

The paper is organized as follows. In Section \ref{sect-model} we
introduce the Heston-Hull-White model. Then in Section \ref{sect-tree} we construct a recombining binomial tree approximation for the pair given by the volatility and the interest rate process. Section \ref{sect-pde} refers to the approximation of functions of the underlying asset price process by means of PDE arguments. In Section \ref{sect-alg} we describe the hybrid tree/finite-difference scheme for the computation of American options. In Section \ref{sect-model2d} we see how to generalize the previous procedure in order to handle the Heston-Hull-White2d process. In Section \ref{sect-MC} we show that our arguments can be used also to set-up simulations, to be applied to construct Monte Carlo algorithms. Finally, numerical results and comparisons with other existing methods are given in Section \ref{sect-numerics}, showing the efficiency of the proposed methods in terms of the results and of the computational time costs.

\section{The Heston-Hull-White model}\label{sect-model}
The Heston Hull-White model concerns with cases where the
volatility $V$ and the interest rate $r$ are assumed to be
stochastic. The dynamics under the risk neutral measure of the share price $S$ and the volatility process $V$ are governed by the stochastic differential equation system
\begin{align*}
&\frac{dS_t}{S_t}= (r_t-\eta) dt+\sqrt{V_t}\, dZ_t,  \\
&dV_t= \kappa_V(\theta_V-V_t)dt+\sigma_V\sqrt{V_t}\,dW^1_t,\\
&dr_t= \kappa_r(\theta_r(t)-r_t)dt+\sigma_r dW^2_t,
\end{align*}
with initial data  $S_0>0$, $V_0>0$ and $r_0>0$, where $Z$, $W^1$ and $W^2$ are suitable and possibly correlated Brownian motions. Recall that $V_t$ is a CIR process whereas $r_t$ is a generalized Ornstein-Uhlenbeck (hereafter OU) process: here $\theta_r$ is not constant but it is a deterministic function which is completely determined by the market values of the zero-coupon bonds (see \cite{bm}).

Let us fix the correlations among the Brownian motions. As observed in \cite{GO}, the important correlations are between the pairs $(S,V)$ and $(S,r)$. So, we assume that $W=(W^1,W^2)$ is a standard Brownian motion in $\R^2$ and $Z$ is a Brownian motion in $\R$ which is correlated both with $W^1$ and  $W^2$:
$$
\mbox{$d\<Z,
W_1\>_t=\rho_1\,dt$ and $d\<Z,
W_2\>_t=\rho_2\,dt$.}
$$
By passing to the logarithm $Y=\ln S$ in the first component and taking into account the above mentioned correlations, we reduce to the dynamics
\begin{align*}
&dY_t=(r_t-\eta-\frac 12 V_t)dt+\sqrt{V_t}\, \big(\rho_1dW^1_t+\rho_2dW^2_t+\rho_3dW^3_t\big), \quad Y_0=\ln S_0\in\R, \\
&dV_t= \kappa_V(\theta_V-V_t)dt+\sigma_V\sqrt{V_t}\,dW^1_t,
\quad V_0>0,\\
&dr_t= \kappa_r(\theta_r(t)-r_t)dt+\sigma_r dW^2_t,
\quad r_0>0,
\end{align*}
where $W=(W^1,W^2,W^3)$ is a standard Brownian motion in $\R^3$ and the correlation parameter $\rho_3$ is given by
{$$
\rho_3=\sqrt{1-\rho_1^2-\rho_2^2}\quad\mbox{with}\quad \rho_1^2+\rho_2^2<1.
$$
}

As already done in \cite{hw1}, the process $r$ can be written in the following way:
\begin{equation}\label{rX}
r_t   = \sigma_r X_t + \varphi_t
\end{equation}
where
\begin{equation}\label{Xphi}
X_t  = - \kappa_r \int_0^tX_s\,ds + \,W^2_t \quad\mbox{and}\quad
\varphi_t=r_0e^{-\kappa_r t}+\kappa_r\int_0^t\theta_r(s)e^{-\kappa_r(t-s)}ds.
\end{equation}
So, we can consider the triple $(Y,V,X)$, whose dynamics is given by
\begin{equation}\label{YVX-dyn}
\begin{array}{ll}
&dY_t=\mu_Y(V_t,X_t, t)dt+\sqrt{V_t}\, \big(\rho_1dW^1_t+\rho_2dW^2_t+\rho_3dW^3_t\big), \quad Y_0=\ln S_0\in\R, \smallskip\\
&dV_t= \mu_V(V_t)dt+\sigma_V\sqrt{V_t}\,dW^1_t,
\quad V_0>0,\smallskip\\
&dX_t=\mu_X(X_t)dt+dW^2_t,
\quad X_0=0,
\end{array}
\end{equation}
where
\begin{align}
\label{muY}
&\mu_Y(v,x,t)=\sigma_rx+\varphi_t-\eta-\frac 12 \,v,\\
\label{muV}
&\mu_V(v)=\kappa_V(\theta_V-v),\\
\label{muX}
&\mu_X(x)= -\kappa_r x.
\end{align}

The purpose of this paper is to efficiently approximate the process $(Y,V,X)$ in order to numerically compute the price of options written on the share process $S$.

\section{The recombining binomial tree for the pair $X$ and $V$}\label{sect-tree}
First of all, we consider an approximation for the pair $(V,X)$ on the
time-interval $[0,T]$ by means of a $2$-dimensional computationally simple tree, that is by means of a Markov chain
that runs over a $2$-dimensional recombining bivariate lattice (recombining binomial tree). In the usual case, as in the Cox-Ross-Rubinstein tree \cite{crr}, at each time step the process can jump either on the nearest up-node or on the nearest down-node. Here, we consider the possibility of ``multiple jumps'' as introduced in Nelson and  Ramaswamy \cite{nr}. Roughly speaking, the process can again jump upward or downward but the up/down jump nodes might not be the nearest ones: they are defined as the up/down positions at the next time-step whose associated transition probabilities better interpolate the theoretical expectation of the transition. As discussed in Nelson and  Ramaswamy \cite{nr}, this is the best way to construct an efficient tree for the approximation of one-dimensional diffusion processes, especially when the diffusion coefficient is not constant. Figure \ref{fig:tree} shows an example of possible ``multiple jumps'' for the trees that approximate our processes $X$ and $V$, that we are going to describe.

\smallskip

In this section, we consider a discretization of the time-interval $[0,T]$ in $N$ subintervals $[nh,(n+1)h]$, $n=0,1,\ldots,N$, with $h=T/N$.

\subsection{The tree for $X$}\label{sect-treeX}
The construction of the recombining binomial tree for the process $X$ is quite standard, because here the diffusion coefficient is constant. For $n=0,1,\ldots,N$, consider the lattice for the process $X$
\begin{equation}\label{state-space-X}
\mathcal{X}_n^h=\{x_{n,j}\}_{j=0,1,\ldots,n}\quad\mbox{with}\quad
x_{n,j}=(2j-n)\sqrt{h}
\end{equation}
(notice that $x_{0,0}=0=X_0$). For each fixed $x_{n,j}\in\mathcal{X}_n^h$, we define the ``up'' and ``down''  jump by means of $j_u^h(n,j)$ and $j_d^h(n,j)$  defined by
\begin{align}
\label{ju}
&j_u^h(n,j) =\min\{j^*\,:\, j+1\leq j^*\leq n+1\mbox{ and }x_{n,j}+\mu_X(x_{n,j})h \le x_{n+1, j^*}\},\\
\label{jd}
&j_d^h(n,j) =\max\{j^*\,:\, 0\leq j^*\leq j \mbox{ and }x_{n,j}+\mu_X(x_{n,j})h \ge x_{n+1, j^*}\},
\end{align}
$\mu_X$ being the drift of the process $X$, see \eqref{muX}.
As usual, one sets $j_u^h(n,j)=n+1$ if $\{j^*\,:\, j+1\leq j^*\leq n+1\mbox{ and }x_{n,j}+\mu_X(x_{n,j})h \le x_{n+1, j^*}\}=\emptyset$ and $j_d^h(n,j)=0$ if $\{j^*\,:\, 0\leq j^*\leq j \mbox{ and }x_{n,j}+\mu_X(x_{n,j})h \ge x_{n+1, j^*}\}=\emptyset$.
Note that the up/down jumps in \eqref{ju}-\eqref{jd} might not be the nearest up/down positions in the lattice at time $n+1$. An example is given in Figure \ref{fig:tree}-left, where the lattice $\mathcal{X}_n^h$ is drawn and some possible instances of $x_{n,j}$, $x_{n+1,j_d^h(n,j)}$ and $x_{n+1,j_u^h(n,j)}$ are shown to exhibit as the tree can be visited.

The transition probabilities are defined in order to better interpolate the expected local transition:  starting from the node $(n,j)$, the probability that the process jumps to $j_u^h(n,j)$ and $j_d^h(n,j)$ at time-step $n+1$ are set as
\begin{equation}\label{pij}
p^{X,h}_u(n,j)
=0\vee \frac{\mu_X(x_{n,j})h+ x_{n,j}-x_{n+1,j_d^h(n,j)} }{x_{n+1,j_u^h(n,j)}-x_{n+1,j_d^h(n,j)}}\wedge 1
\quad\mbox{and}\quad p^{X,h}_d(n,j)=1-p^{X,h}_u(n,j)
\end{equation}
respectively. This gives rise to a Markov chain $(\hat X^h_{n})_{n=0,\ldots,N}$ that
weakly converges, as $h\to 0$, to the diffusion process
$(X_{t})_{t\in[0,T]}$ and turns out to be a robust tree approximation
for the OU process $X$.

\subsection{The tree for $V$}\label{sect-treeV}

For the CIR volatility process $V$, we consider a recombining binomial tree procedure that again follows the ``multiple jumps'' approach. In this case the recombining lattice is built by means of the transformation 
$$
f(V_t)=\frac 2{\sigma_V}\,\sqrt{V_t}.
$$
This transformation is particularly important because $f(V_t)$ turns
out to be a diffusion process with unit diffusion coefficient, and
this fact is useful in order to construct a recombining lattice. Many
authors (see e.g. \cite{hst} or \cite{wei}) propose tree algorithms
for $V_t$ by working on  the transformed process $f(V_t)$. The unpleasant fact is that now the drift of $f(V_t)$ is very bad and is such that the approximating process converges only when the Feller condition holds: $2\kappa_V\theta_V\geq \sigma^2_V$. In order to overcome this fact, we use the approach in \cite{bib:acz}, that, roughly speaking, works as follows:  the tree structure is built by using again $f$ (see next \eqref{state-space-V}) but the possible jumps and the transition probabilities are set on the dynamics  of the original (and not transformed) CIR process $V_t$ (see next \eqref{ku}-\eqref{kd} and
\eqref{pik}). The main fact is that now the weak convergence on the path space is achieved for every values of $\kappa_V,\theta_V,\sigma_V>0$, so the Feller condition is not required. Details and comparisons with other tree existing methods to approximate the CIR process are given in \cite{bib:acz}.

\smallskip

\begin{figure}[htp]

  \centering

\begin{tabular}{cc}

\includegraphics[scale=0.5]{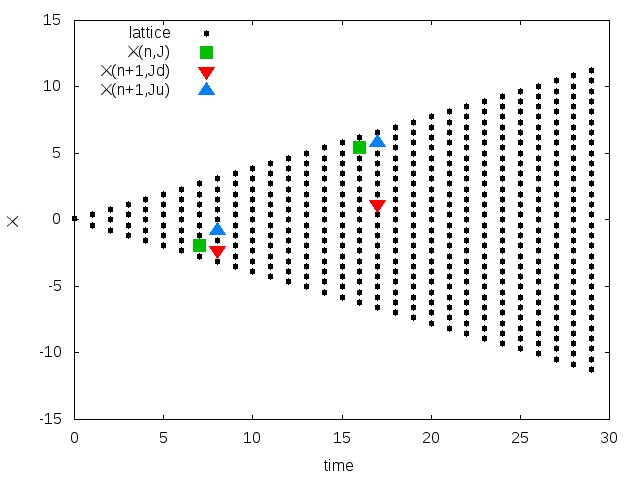}

&

\includegraphics[scale=0.5]{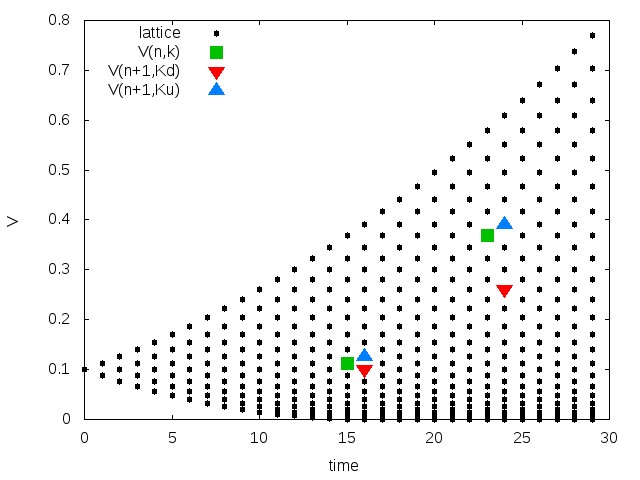}

\end{tabular}
\caption{Example of a tree for $X$ on the left and of a tree for $V$ on the right.}
\label{fig:tree}

\end{figure}

For $n=0,1,\ldots,N$, consider the lattice
\begin{equation}\label{state-space-V}
\mathcal{V}_n^h=\{v_{n,k}\}_{k=0,1,\ldots,n}\quad\mbox{with}\quad
v_{n,k}=\Big(\sqrt {V_0}+\frac{\sigma_V} 2(2k-n)\sqrt{h}\Big)^2\I_{\sqrt {V_0}+\frac\sigma 2(2k-n)\sqrt{h}>0}
\end{equation}
(notice that $v_{0,0}=V_0$). For each fixed $v_{n,k}\in\mathcal{V}_n^h$, we define the ``up'' and ``down'' jump by means of
\begin{align}
\label{ku}
&k_u^h(n,k) =\min\{k^*\,:\, k+1\leq k^*\leq n+1\mbox{ and }v_{n,k}+\mu_V(v_{n,k})h \le v_{n+1, k^*}\},\\
\label{kd}
&k_d^h(n,k) =\max\{k^*\,:\, 0\leq k^*\leq k \mbox{ and }v_{n,k}+\mu_V(v_{n,k})h \ge v_{n+1, k^*}\}
\end{align}
where the drift $\mu_V$ of $V$ is defined in \eqref{muX} and
with the understanding $k_u^h(n,k)=n+1$ if $\{k^*\,:\, k+1\leq k^*\leq n+1\mbox{ and }v_{n,k}+\mu_V(v_{n,k})h \le v_{n+1, k^*}\}=\emptyset$ and $k_d^h(n,k)=0$ if $\{k^*\,:\, 0\leq k^*\leq k \mbox{ and }v_{n,k}+\mu_V(v_{n,k})h \ge v_{n+1, k^*}\}=\emptyset$.
By construction, the up/down jumps in \eqref{ku}-\eqref{kd} might not be the nearest up/down positions in the lattice at time $n+1$. In Figure \ref{fig:tree}-right we show an example of the lattice $\mathcal{V}_n^h$ together with some possible instances of the triple $(v_{n,k},v_{n+1,k_d^h(n,j)},v_{n+1,k_u^h(n,j)})$.

The transition probabilities are defined in order to better interpolate the expected local transition: starting from the node $(n,k)$ the probability that the process jumps to $k_u^h(n,k)$ and $k_d^h(n,k)$ at time-step $n+1$ are set as
\begin{equation}\label{pik}
p^{V,h}_u(n,k)
=0\vee \frac{\mu_V(v_{n,k})h+ v_{n,k}-v_{n+1,k_d^h(n,k)} }{v_{n+1,k_u^h(n,k)}-v_{n+1,k_d^h(n,k)}}\wedge 1
\quad\mbox{and}\quad p^{V,h}_d(n,k)=1-p^{V,h}_u(n,k)
\end{equation}
respectively. This gives rise to a Markov chain $(\hat V^h_{n})_{n=0,\ldots,N}$ that weakly converges, as $h\to 0$, to the diffusion process $(V_{t})_{t\in[0,T]}$ and turns out to be a robust tree approximation for the CIR process $V$ - details are given in \cite{bib:acz}.

\subsection{The tree for the pair $(V,X)$}\label{sec:treeVX}

The tree procedure for the pair $(V,X)$ is set by joining the trees built for $V$ and for $X$. Namely, for $n=0,1,\ldots,N$, consider the lattice
\begin{equation}\label{state-space-Vr}
\mathcal{V}_n^h\times \mathcal{X}_n^h=\{(v_{n,k},x_{n,j})\}_{k,j=0,1,\ldots,n}.
\end{equation}
Starting from the node $(n,k,j)$, which corresponds to the position $(v_{n,k},x_{n,j})\in\mathcal{V}_n^h\times \mathcal{X}_n^h$, we define the four possible jump by setting the four nodes at time $n+1$ following the definitions \eqref{ju}-\eqref{jd} and \eqref{ku}-\eqref{kd}:
\begin{equation}\label{treescheme}
\begin{array}{lcr}
(n+1,k_u^h(n,k),j_u^h(n,j)) & \hbox{with probability} & p^{h}_{uu}(n,k,j)=p^{V,h}_{u}(n,k)p^{X,h}_{u}(n,j), \smallskip\\
(n+1,k_u^h(n,k),j_d^h(n,j)) & \hbox{with probability} & p^{h}_{ud}(n,k,j)=p^{V,h}_{u}(n,k)p^{X,h}_{d}(n,j), \smallskip\\
(n+1,k_d^h(n,k),j_u^h(n,j)) & \hbox{with probability} & p^{h}_{du}(n,k,j)=p^{V,h}_{d}(n,k)p^{X,h}_{u}(n,j), \smallskip\\
(n+1,k_d^h(n,k),j_d^h(n,j)) & \hbox{with probability} & p^{h}_{dd}(n,k,j)=p^{V,h}_{d}(n,k)p^{X,h}_{d}(n,j),
\end{array}
\end{equation}
where the above probabilities $p^{V,h}_{u}(n,k)$, $p^{V,h}_{d}(n,k)$, $p^{X,h}_{u}(n,j)$ and $p^{X,h}_{d}(n,j)$ are defined in \eqref{pik} and \eqref{pij} respectively. The above factorization is due to the orthogonality of the noises driving the two processes. As a quite immediate consequence of standard results (see e.g. the techniques in \cite{nr}), one gets the following: the associated bivariate Markov chain $(\hat V^h_n,\hat X^h_n)_{n=0,\ldots,N}$ weakly converges to the diffusion pair $(V_t,X_t)_{t\in[0,T]}$ solution to
\begin{align*}
&dV_t= \mu_V(V_t)dt+\sigma_V\sqrt{V_t}\,dW^1_t,
\quad V_0>0,\\
&dX_t  = - \kappa_r X_t\,dt + \sigma_r\,dW^2_t,\quad X_0=0.
\end{align*}

\begin{remark}\label{corr-VX}
In the case one is interested in introducing a correlation between the noises $W^1$ and $W^2$ driving the process $V$ and $X$ respectively, the joint tree can be constructed on the same lattice but the jump probabilities are no more of a product-type: the transition
probabilities $p^{h}_{uu}(n,k,j)$, $p^{h}_{ud}(n,k,j)$, $p^{h}_{du}(n,k,j)$ and $p^{h}_{dd}(n,k,j)$  can be
computed by matching (at the first order in $h$) the conditional mean and the
conditional covariance between the continuous and the discrete processes of $V$
and $X$. More precisely, for both components the conditional mean is matched by construction (this is actually the main consequence of the definition of the multiple jumps). As for the conditional covariance, assuming that $d\<W^1,W^2\>_t=\alpha dt$, with $|\alpha|<1$, then  one has $d\<V,X\>_t=\alpha\sigma_V\sqrt{V_t}\,dt$. Therefore, the matching conditions lead to solving the following system:
$$
\left\{
\begin{array}{l}
p^{h}_{uu}(n,k,j)+p^{h}_{ud}(n,k,j)=p^{V,h}_{u}(n,k)\smallskip\\
p^{h}_{uu}(n,k,j)+p^{h}_{du}(n,k,j)=p^{X,h}_{u}(n,j)\smallskip\\
p^{h}_{uu}(n,k,j)+p^{h}_{ud}(n,k,j)+
p^{h}_{du}(n,k,j)+p^{h}_{dd}(n,k,j)=1\smallskip\\
m^h_{uu}(n,k,j)p^{h}_{uu}(n,k,j)
+m^h_{ud}(n,k,j)p^{h}_{ud}(n,k,j)+\smallskip\\
\quad+m^h_{du}(n,k,j)p^{h}_{du}(n,k,j)
+m^h_{dd}(n,k,j)p^{h}_{dd}(n,k,j)
=\alpha \sigma_V \sqrt{v_{n,k}}\,h
\end{array}
\right.
$$
where
$$
\begin{array}{ll}
&m^h_{uu}(n,k,j)=(v_{n+1,k^h_u(n,k)}-v_{n,k})(x_{n+1,j^h_u(n,j)}-x_{n,j}),\smallskip\\
&m^h_{ud}(n,k,j)=(v_{n+1,k^h_u(n,k)}-v_{n,k})(x_{n+1,j^h_d(n,j)}-x_{n,j}),\smallskip\\
&m^h_{du}(n,k,j)=(v_{n+1,k^h_d(n,k)}-v_{n,k})(x_{n+1,j^h_u(n,j)}-x_{n,j}),\smallskip\\
&m^h_{dd}(n,k,j)=(v_{n+1,k^h_d(n,k)}-v_{n,k})(x_{n+1,j^h_d(n,j)}-x_{n,j}).
\end{array}
$$
This is done in \cite{bib:acz} in a different context but the proof of the weak convergence on the path space is analogous - this can be done by standard arguments, as in \cite{nr} or \cite{ek}.
\end{remark}

\section{Approximating the $Y$-component: the finite-difference approach}\label{sect-pde}

We go now back to \eqref{YVX-dyn}, that is
$$
\begin{array}{ll}
&dY_t=\mu_Y(V_t,X_t,t)dt+\sqrt{V_t}\, \big(\rho_1dW^1_t+\rho_2dW^2_t+\rho_3dW^3_t\big), \quad Y_0=\ln S_0, \smallskip\\
&dV_t= \mu_V(V_t)dt+\sigma_V\sqrt{V_t}\,dW^1_t,
\quad V_0>0,\smallskip\\
&dX_t=\mu_X(X_t)dt+dW^2_t,
\quad X_0=0,
\end{array}
$$
where $\mu_Y$, $\mu_V$ and $\mu_X$ are given in \eqref{muY}, \eqref{muV} and \eqref{muX} respectively. By isolating $\sqrt{V_t}dW^1_t$ in the second line and $dW^2_t$ in the third one, we obtain
\begin{equation}\label{Y1}
dY_t=\frac{\rho_1}{\sigma_V}dV_t+\rho_2\sqrt{V_t}dX_t+\mu(V_t,X_t,t)dt+\rho_3\sqrt{V_t}\,dW^3_t \end{equation}
with
\begin{equation}\label{mu-fin}
\begin{array}{rl}
\mu(v,x,t)
&=\mu_Y(v,x,t)-\frac{\rho_1}{\sigma_V}\mu_V(v)-\rho_2\sqrt{v}\,\mu_X(x)\smallskip\\
&=\sigma_rx+\varphi_t-\eta-\frac 12 \,v
-\frac{\rho_1}{\sigma_V}\,\kappa_V(\theta_V-v)+\rho_2\kappa_r x\sqrt v.
\end{array}
\end{equation}
What we are going to do is mainly based on the fact that the noise $W^3$ is independent of the processes $V$ and $X$.

\subsection{The approximating scheme for the triple $(Y,V,X)$}\label{sect-apprYVX}

We consider an approximating process $Y^h$ for $Y$ turning out by freezing the coefficients in \eqref{Y1}: we define $Y^h_0=Y_0$ and for $t\in[nh,(n+1)h]$ with $n=0,1,\ldots,N-1$ we set
\begin{align*}
Y^h_t
=&Y^h_{nh}+
\frac{\rho_1}{\sigma_V}(V_t-V_{nh})
+\rho_2\sqrt{V_{nh}}(X_t-X_{nh})+\mu(V_{nh},X_{nh},{nh})(t-nh)+\rho_3\sqrt{V_{nh}}\,(W^3_t-W^3_{nh}).
\end{align*}

We consider now the approximating tree $(\hat V^h_n,\hat X^h_n)_{n\in\{0,\ldots,N\}}$ and we call $(\bar V^h_t,\bar X^h_t)_{t\in[0,T]}$ the associated time-continuous approximating process for the pair $(V,X)$, that is
$$
\bar V^h_t=\hat V^h_{\lfloor t/h\rfloor}\quad\mbox{and}\quad
\bar X^h_t=\hat X^h_{\lfloor t/h\rfloor}.
$$
We then assume that the noise driving the pair $(\bar V^h_t,\bar X^h_t)_{t\in[0,T]}$ is independent of the Brownian motion $W^3$ and we insert this discretization for $(V,X)$ in the discretization scheme for $Y$. So, we obtain our final approximating process $\bar Y^h_t$ by setting
$\bar Y^h_0=Y_0$ and for $t\in[nh,(n+1)h]$ with $n=0,1,\ldots,N-1$ then
\begin{equation}\label{barYh}
\bar Y^h_t
=Y^h_{nh}+
\frac{\rho_1}{\sigma_V}(\bar V^h_t-\bar V^h_{nh})
+\rho_2\sqrt{\bar V^h_{nh}}(\bar X^h_t-\bar X^h_{nh})+\mu(\bar X^h_{nh},\bar V^h_{nh},{nh})(t-nh)+\rho_3\sqrt{\bar V^h_{nh}}\,(W^3_t-W^3_{nh}).
\end{equation}
Notice that if we set
\begin{equation}\label{Z}
\bar Z^{h}_t=\bar Y^h_t
-\frac{\rho_1}{\sigma_V}(\bar V^h_t-\bar V^h_{nh})
-\rho_2\sqrt{\bar V^h_{nh}}(\bar X^h_t-\bar X_{nh}), \quad t\in [nh,(n+1)h]
\end{equation}
then we have
\begin{equation}\label{barZ}
\begin{array}{l}
d\bar Z^{h}_t=\mu(\bar X^h_{nh},\bar V^h_{nh},{nh})dt+\rho_3\sqrt{\bar V^h_{nh}}\,dW^3_t,\quad t\in (nh,(n+1)h],\smallskip\\
\bar Z^{h}_{nh}=\bar Y^h_{nh}
\end{array}
\end{equation}
that is $\bar Z^h$ solves a SDE with constant coefficients and at time $nh$ it starts from $\bar Y^h_{nh}$. Take now a function $f$: we are interested in approximating
$$
\E(f(Y_{(n+1)h})\mid Y_{nh}=y, V_{nh}=v, X_{nh}=x).
$$
By using our scheme and the process $\bar Z^h$ in \eqref{Z}, we approximate it with the expectation done on the approximating processes, that is
$$
\begin{array}{l}
\displaystyle
\E\big(f(\bar Y^h_{(n+1)h})\mid \bar Y^h_{nh}=y, \bar V^h_{nh}=v, \bar X^h_{nh}=x\big)\smallskip\\
\displaystyle
\qquad=\E\big(f(\bar Z^{h}_{(n+1)h}+\frac{\rho_1}{\sigma_V}(\bar V^h_{(n+1)h}-\bar V^h_{nh})
+\rho_2\sqrt{\bar V^h_{nh}}(\bar X^h_{(n+1)h}-\bar X^h_{nh})
)\mid \bar Z^{h}_{nh}=y, \bar V^h_{nh}=v, \bar X^h_{nh}=x\big).
\end{array}
$$
Since $(\bar V^h, \bar X^h)$ is independent of the Brownian noise $W^3$ driving $\bar Z^h$ in \eqref{Z}, we can write
\begin{equation}\label{E1}
\begin{array}{l}
\displaystyle
\E(f(\bar Y^h_{(n+1)h})\mid \bar Y^h_{nh}=y, \bar V^h_{nh}=v, \bar X^h_{nh}=x)\smallskip\\
\qquad=\E\Big(\Psi_f\Big(\frac{\rho_1}{\sigma_V}(\bar V^h_{(n+1)h}-v)
+\rho_2\sqrt{v}(\bar X^h_{(n+1)h}-x); y,v,x
\Big)\,\Big|\, \bar V^h_{nh}=v, \bar X^h_{nh}=x\Big),
\end{array}
\end{equation}
in which
\begin{equation}\label{E2}
\Psi_f(\xi; y,v,x)=\E(f(\bar Z^h_{(n+1)h}+\xi
)\mid \bar Z^h_{nh}=y, \bar V^h_{nh}=v, \bar X^h_{nh}=x).
\end{equation}
Now, in order to compute the above quantity $\Psi_f(\xi)$, consider a generic function $g$ and set
$$
u(s,z;v,x)=\E(g(\bar Z^h_{(n+1)h})\mid \bar Z^h_{s}=z, \bar V^h_{s}=v, \bar X^h_{s}=x),\quad s\in[nh,(n+1)h].
$$
By \eqref{barZ} and the Feynman-Kac representation formula we can state that, for every fixed $x\in\R$ and $v\geq 0$, the function $(s,z)\mapsto u(s,z;v,x)$ is the solution to
\begin{equation}\label{uPDE}
\left\{\begin{array}{l}
\partial_s u+\mu(v,x,s)\partial_z u+\frac 12 \rho_3^2v\partial^2_{z}u=0,\quad s\in[nh,(n+1)h), \ z\in\R, \smallskip\\
u((n+1)h,z;v,x)=g(z),
\end{array}\right.
\end{equation}
$\mu$ being given in \eqref{mu-fin}. In order to solve the PDE problem \eqref{uPDE}, we use a finite-difference approach.

\subsection{Finite-differences}\label{sect-FD}

At each time step $n$ we numerically solve \eqref{uPDE} at time $s=nh$ by applying finite-difference techniques.

We fix a grid on the $y$-axis $\mathcal{Y}_M=\{y_i=Y_0+i\Dy\}_{i\in\mathcal{J}_{M}}$, with $\mathcal{J}_{M}=\{-M,\ldots,M\}$ and $\Delta y=y_i-y_{i-1}$. For fixed $n$, $v\geq 0$ and $x\in\R$, we set $u^n_i=u(nh, y_i;v,x)$ the discrete solution of \eqref{uPDE} at time $nh$ on the point $y_i$ of the grid $\mathcal{Y}_{M}$ - for simplicity of notations, we do not stress in $u^n_i$ the dependence on $v$ and $x$ (from the coefficients of the PDE).

The finite difference method we are going to set is inspired from the one developed in \cite{bcz}. But a main difference arises: here, we do not distinguish anymore between the diffusion dominant or reaction dominant case and we propose to apply a full implicit finite-difference approximation in time. In fact,
the discrete solution $u^n$ to problem \eqref{uPDE} at time $nh$ is computed in terms of the solution $u^{n+1}$ at time $(n+1)h$ by using the following finite-difference scheme:
\begin{equation}\label{discr_eq}
\begin{array}{l}
\Frac{u^{n+1}_i-u^n_i}{h}+\mu(v,x,nh)\Frac{u^{n}_{i+1}-u^{n}_{i-1}}{2\Dy}+\frac{1}{2}\rho_3^2\ v\ \Frac{u^{n}_{i+1}-2u^{n}_i+u^{n}_{i-1}}{\Dy^2}=0.
\end{array}
\end{equation}
Of course, \eqref{discr_eq} has to be coupled with suitable numerical boundary relations.
We assume that the boundary values are defined by the following Neumann-type conditions:
\begin{equation}
u^{n}_{-M-1}=u^{n}_{-M+1},
\quad u^{n}_{M+1}=u^{n}_{M-1}.\label{num_boundary_cond_impl}
\end{equation}

Then, by applying the implicit finite-difference \eqref{discr_eq} coupled with the boundary conditions \eqref{num_boundary_cond_impl}, we get the solution $u^n=(u^n_{-M},\ldots,u^n_M)^T$ by solving the following linear system
\begin{equation}\label{lin_sys}
A\, u^n = u^{n+1},
\end{equation}
where $A=A(v,x)$ is the $(2M+1)\times (2M+1)$ tridiagonal real matrix given by
\begin{equation}\label{matrixA}
A = \left(
\begin{array}{ccccc}
1+2\beta & -2\beta & & & \\
-\beta+\alpha & 1+2\beta & -\beta-\alpha & & \\
 & \ddots & \ddots & \ddots &  \\
 &   & -\beta+\alpha & 1+2\beta & -\beta-\alpha\\
 &   &        &  -2\beta  & 1+2\beta
\end{array}
\right),
\end{equation}
with
\begin{equation}\label{alphabeta_impl}
\alpha=\frac{h}{2\Dy}\,\mu(v,x,nh)\quad\mbox{and}\quad\beta=\frac{h}{2\Dy^2}\,\rho_3^2v,
\end{equation}
$\mu$ being defined in \eqref{mu-fin}.
We stress that at each time step $n$, the quantities $v$ and $x$ are constant and known values (defined by the tree procedure for the pair $(V,X)$) and then $\alpha$ and $\beta$ are constant parameters too.

{One can easily see that the implicit scheme \eqref{discr_eq} is unconditionally stable. Moreover, by applying standard results (Theorem 2.1 in \cite{bt} e.g.), the matrix $A$ is invertible for $\beta\neq|\alpha|$. Therefore, setting
 \begin{equation}\label{pih}
\Pi(v,x)= A^{-1}(v,x),
\end{equation}
the numerical solution to \eqref{uPDE} on the grid $\mathcal{Y}_{M}$ through the above discretization procedure is given by
\begin{equation}\label{FD}
u(nh,y_i;v,x)\simeq u^n_i
=\sum_{\ell\in\mathcal{J}_M}\Pi_{i\ell}(v,x)g(z_\ell),\quad i\in \mathcal{J}_{M}.
\end{equation}

\begin{remark}\label{rem-boundary}
Other numerical boundary conditions can surely be selected, for example the two \emph{boundary} values $u^{n}_{-M}$ and $u^{n}_{M}$ may be a priori fixed by a known constant (this procedure typically appears in financial problems).
\end{remark}

\subsection{The scheme on the $Y$-component}

We can now come back to our original problem, that is the computation of the function $\Psi_f(\xi;y,v,x)$ in \eqref{E2} allowing one to numerically compute the expectation in \eqref{E1}.

We consider the approximating process $(\bar Y^h,\bar V^h,\bar X^h)$ as described in Section \ref{sect-apprYVX}. This means that the pair $(v,x)$ at time-step $n$ is located on the lattice $\mathcal{V}_n^h\times \mathcal{X}_n^h$: $v=v_{n,k}$ and $x=x_{n,j}$, for $0\leq k,j\leq n$. Then \eqref{FD} gives the following approximation: for each $y_{i}\in \mathcal{Y}_{M}$,
$$
\Psi_f\big(\xi;y_i,v_{n,k},x_{n,j}\big)
\simeq
\sum_{\ell\in\mathcal{J}_M}\Pi_{i\ell}(v_{n,k},x_{n,j})f\big(y_{\ell}+\xi\big), \quad i\in\mathcal{J}_M.
$$
Therefore, the expectation in \eqref{E1} is computed on the approximating tree for $(V,X)$ by means of the above approximation:
\def\T{\mathrm{T}}
\begin{equation}\label{E3}
\E(f(\bar Y^h_{(n+1)h})\mid \bar Y^h_{nh}=y_i, \bar V^h_{nh}=v_{n,k}, \bar X^h_{nh}=x_{n,j})\simeq
\!\!\!\sum_{a,b\in\{d,u\}}
\sum_{\ell\in\mathcal{J}_M}\Pi_{i\ell}(v_{n,k},x_{n,j})\T_{n,k,j}f(\ell,a,b)
p^{h}_{ab}(n,k,j)
\end{equation}
where
$$
\T_{n,k,j}f(\ell,a,b)
=f\Big(y_{\ell}
+\frac{\rho_1}{\sigma_V}(v_{n+1,k_a(n,k)}-v)+\rho_2\sqrt{v}(x_{n+1,j_b(n,j)}-x)\Big)
$$
and the jump probabilities $p^{h}_{ab}(n,k,j)$ are given in \eqref{treescheme} (or in Remark \ref{corr-VX} if a correlation is assumed between the noises driving $V$ and $X$).

Similar arguments can be used in order to compute the conditional expectation in the left hand side of \eqref{E3} when the function $f$ depends on the variables $v$ and $x$ also. Then one gets
\begin{equation}\label{E4}
\begin{array}{l}
\displaystyle
\E(f(\bar Y^h_{(n+1)h},\bar V^h_{(n+1)h}, \bar X^h_{(n+1)h},)\mid \bar Y^h_{nh}=y_i, \bar V^h_{nh}=v_{n,k}, \bar X^h_{nh}=x_{n,j})\smallskip\\
\qquad\simeq\displaystyle
\sum_{a,b\in\{d,u\}}
\sum_{\ell\in\mathcal{J}_M}\Pi_{i\ell}(v_{n,k},x_{n,j})\T_{n,k,j}f(\ell,a,b)
p^{h}_{ab}(n,k,j)
\end{array}
\end{equation}
where
\begin{equation}\label{T}
\begin{array}{l}
\displaystyle\T_{n,k,j}f(\ell,a,b)=\smallskip\\
\displaystyle\quad=f\Big(y_{\ell}
+\frac{\rho_1}{\sigma_V}(v_{n+1,k_a(n,k)}-v_{n,k})+\rho_2\sqrt{v_{n,k}}(x_{n+1,j_b(n,j)}-x_{n,j}), v_{n+1,k_a(n,k)}, x_{n+1,j_b(n,j)}\Big).
\end{array}
\end{equation}

\section{The algorithm for the pricing of  American options}\label{sect-alg}

The natural application of the hybrid tree/finite-difference approach arises in the pricing of American options. Consider an American option with maturity $T$ and payoff function $(\Phi(S_t))_{t\in[0,T]}$. First of all, we consider the log-price process, so the obstacle will be given by
$$
\Psi(Y_t)=\Phi(e^{Y_t}),\quad t\in[0,T].
$$
The price $P(t,y,v,x)$ of such an American option is given by (recall the relation between the interest rate $r$ and the process $X$: $r_t=\sigma_rX_t+\varphi_t$, see \eqref{rX})
$$
P(t,y,v,x)
=\sup_{\tau\in \mathcal{T}_{t,T}}\E\Big(e^{-\int_t^\tau (\sigma_rX^{t,x}_s+\varphi_s)ds}\Psi(Y^{t,y,v,x}_\tau)\Big)
$$
where $\mathcal{T}_{t,T}$ denotes the set of all stopping times taking values on $[t,T]$. Hereafter,  $(Y^{t,y,v,x},V^{t,v},X^{t,x})$ denotes the solution of the SDE \eqref{YVX-dyn} starting at $(y,v,x)$ at time $t$.

The price at time $0$ of such an option
is then approximated by a backward dynamic programming algorithm. Consider a  discretization of the time interval $[0,T]$ into $N$ subintervals of length $h=T/N$: $[0,T]=\cup_{n=0}^{N-1}[nh,(n+1)h]$.
Then $P(0,Y_0,V_0,X_0)$ is numerically approximated through the quantity $P_h(0,Y_0,V_0,X_0)$ which is iteratively defined as follows: for $(y,v,x)\in\R\times\R_+\times \R$, 
\begin{equation*}\label{backward}
  \begin{cases}
    P_h(T,y,v,x)= \Psi(y)\quad \mbox{and as $n=N-1,\ldots,0$}\\
   P_h(nh,y,v,x) =  \max \Big\{\Psi(y), e^{-(\sigma_r x + \varphi_{nh})h}
   \E\Big(P_h\big((n+1)h, Y^{nh,y,v,x}_{(n+1)h}, V^{nh,v}_{(n+1)h}, X^{nh,x}_{(n+1)h}\big)
   \Big)\Big\}.
  \end{cases}
\end{equation*}
From the financial point of view, this means to allow the exercise at the fixed dates $nh$, $n=0,\ldots,N$.

Consider now the discretization scheme $(\bar Y^h,\bar V^h,\bar X^h)$ discussed in Section \ref{sect-pde}. We use the approximation \eqref{E4} for the conditional expectations that have to be computed at each time step $n$. So, for every point $(y_i,v_{n,k}, x_{n,j})\in \mathcal{Y}_M\times\mathcal{V}_n^h\times\mathcal{X}_n^h$, \eqref{E4} gives
\begin{equation}\label{E5}
\begin{array}{l}
\displaystyle
\E\Big(P_h\big((n+1)h, Y_{(n+1)h}^{nh,y_{i},v_{n,k},x_{n,j}}, V_{(n+1)h}^{nh,v_{n,k}}, X_{(n+1)h}^{nh,x_{n,j}}\big)\Big)\\
\quad\simeq\displaystyle
\sum_{a,b\in\{d,u\}}
\sum_{\ell\in\mathcal{J}_M}\Pi_{i\ell}(v_{n,k},x_{n,j})
\S_{n,k,j}P_h(\ell,a,b)\, p^{h}_{ab}(n,k,j)
\end{array}
\end{equation}
where $\S_{n,k,j}P_h$ denotes the operator in \eqref{T} applied to the function $P_h((n+1)h,\cdot)$, that is
\begin{equation}\label{E5bis}
\begin{array}{l}
\displaystyle
\S_{n,k,j}P_h(\ell,a,b)\\
\displaystyle
=P_h\Big((n+1)h,y_{\ell}
+\frac{\rho_1}{\sigma_V}(v_{n+1,k_a(n,k)}-v_{n,k})
+\rho_2\sqrt{v_{n,k}}(x_{n+1,j_b(n,j)}-x_{n,j}),
v_{n+1,k_a(n,k)}, x_{n+1,j_b(n,j)}\Big).
\end{array}
\end{equation}
We finally summarize the backward induction giving our approximating algorithm. For $n=0,1,\ldots,N$, we define $\tilde P_h(nh,y,v,x)$ for $(y,v,x)\in \mathcal{Y}_M\times\mathcal{V}_n^h\times\mathcal{X}_n^h$ as follows:
\begin{equation}\label{backward-ter}
\begin{cases}
\tilde P_h(T,y_i,v_{N,k},x_{N,j})= \Psi(y_i)\quad \mbox{
and as $n=N-1,\ldots,0$:}\\
\displaystyle
\tilde P_h(nh,y_i,v_{n,k},x_{n,j}) =
\max \Big\{\Psi(y_i),
e^{-(\sigma_r x_{n,j} + \varphi_{nh})h}\times
\\
\quad
\qquad\qquad\qquad\qquad\qquad\qquad\qquad
\times\displaystyle
\sum_{a,b\in\{d,u\}}
\sum_{\ell\in\mathcal{J}_M}\Pi_{i\ell}(v_{n,k},x_{n,j})
p^{h}_{ab}(n,k,j)
\S_{n,k,j}\tilde P_h(\ell,a,b)\Big\}.
  \end{cases}
\end{equation}
Notice that, by \eqref{E5bis}, the computation of $\S_{n,k,j}\tilde P_h(\ell,a,b)$ requires the knowledge of the function $y\mapsto \tilde P_h((n+1)h,y,v,x)$ in points $y$'s that do not necessarily belong to the grid $\mathcal{Y}_M$. Therefore, in practice we compute such a function by means of quadratic interpolations.

\begin{remark}\label{weight}
Let us stress that the r.h.s. of \eqref{E5} can be read in two equivalents ways. First, the term
$$
\sum_{\ell\in\mathcal{J}_M}\Pi_{i\ell}(v_{n,k},x_{n,j})
\S_{n,k,j}P_h(\ell,a,b), \quad a,b\in\{d,u\}, i\in \mathcal{Y}_M,
$$
is the numerical solution to the PDE \eqref{uPDE} with final condition as in \eqref{E5bis}, so the r.h.s. of \eqref{E5} is actually a weighted sum of the four solutions from each jump node $(a,b)\in\{d,u\}$ for the pair $(V,X)$, with weights given by the jump probabilities. But since the differential operator is linear in the Cauchy conditions, then one can first do the weighted sum of the final conditions, that is
$$
\sum_{a,b\in\{d,u\}}
\S_{n,k,j}P_h(\ell,a,b)\, p^{h}_{ab}(n,k,j),\quad \ell\in\mathcal{Y}_M,
$$
and then apply the matrix $\Pi(v_{n,k},x_{n,j})$, i.e. solve the PDE \eqref{uPDE} just once, and this is of course  computationally less expensive.
\end{remark}
We can resume the main steps of our algorithm as follows.
\begin{itemize}
\item
\textsc{Preprocessing:}
  \begin{itemize}
  \item
set the lattice $x_{n,j}$, $0\leq j\leq n\leq N$, for the process $X$ by using (\ref{state-space-X});
  \item
set the lattice $v_{n,k}$, $0\leq k\leq n\leq N$, for the process the $V$ by using (\ref{state-space-V});
  \item
merge the above lattices in a bivariate one $(v_{n,k},x_{n,j})$, $0\leq k,j\leq n\leq N$, by using (\ref{state-space-Vr});
\item
compute the jump-nodes and the transition probabilities $p_{ab}$, $(a,b)\in\{d,u\}$, using \eqref{treescheme};
\item
set a mesh grid $y_i$, $i\in\mathcal{Y}_M$, for the solution of all the PDE's.
\end{itemize}
\item \textsc{Step $N$:} 
for each node $(v_{N,k},
  x_{N,j})$, $0\leq k,j\leq N$, compute the option prices at maturity for each $y_i$, $i\in\mathcal{Y}_M$, by using the payoff function.
\item \textsc{Step $n=N-1,\ldots 0$:} 
for each $(v_{n,k},x_{n,j})$, $0\leq k,j\leq n$, compute the option prices for each $y_i$, $i\in \mathcal{Y}_M$, by solving PDE \eqref{uPDE} through \eqref{FD}, with terminal condition given by the weighted sum of the values at nodes $(a,b)\in\{u,d\}$ which have been computed in the previous step - weight by using the transition probabilities $p_{ab}$ (recall Remark \ref{weight}).
\end{itemize}

The theoretical proof of the convergence of our method is postponed to a further study. Although the ideas inspiring the method  mainly come from \cite{bcz}, here the convergence  problem has to be tackled differently. In fact, in \cite{bcz} the numerical scheme is written through a matrix $\Pi$ which is stochastic, so one can link the scheme to a Markov chain that approximates the process $(Y,V,X)$ and use probabilistic methods (weak convergence) in order to study the convergence. But the scheme proposed here is purely numerical: the matrix $\Pi(v,x)= A^{-1}(v,x)$ in \eqref{pih} is stochastic if and only if $\beta<|\alpha|$, so the link with Markov chains fails and the probabilistic weak convergence cannot be used anymore. So, here we restrict ourselves to the study of the behavior and the efficiency of the proposed approach from the numerical point of view, see next Section \ref{sect-numerics}.

\section{Generalization to the Heston-Hull-White2d model}\label{sect-model2d}

The Heston-Hull-White2d  model generalizes the previous model in the
fact that the quantity $\eta$ is assumed to be stochastic and to follow a diffusion model itself. So, the underlying process is now $4$-dimensional and is given by: the share
price $S$, the volatility process $V$, the interest rate $r$ and the
continuous dividend rate $\eta$. Actually, here the process $\eta$ has not necessarily the meaning of a dividend rate, being for example a further interest rate process. In fact, the Heston-Hull-White2d model occurs in multi-currency models with short-rate interest rates, see e.g. \cite{go12}.

Under the risk neutral measure, the dynamics  are governed by the stochastic differential equation
\begin{align*}
&\frac{dS_t}{S_t}= (r_t-\eta_t) dt+\sqrt{V_t}\, dZ_t,  \\
&dV_t= \kappa_V(\theta_V-V_t)dt+\sigma_V\sqrt{V_t}\,dW^1_t,\\
&dr_t= \kappa_r(\theta_r(t)-r_t)dt+\sigma_r dW^2_t,\\
&d\eta_t= \kappa_\eta(\theta_\eta(t)-\eta_t)dt+\sigma_\eta dW^3_t,
\end{align*}
with initial data  $S_0, V_0, r_0,\eta_0>0$, where $Z$, $W^1$, $W^2$
and $W^3$ denote possibly correlated Brownian motions. Note
that the process $\eta$ evolves as a generalized OU
process: $\theta_\eta$ is a deterministic
function of the time.

We consider non null correlations between the Brownian motions driving the pairs $(S,V)$, $(S,r)$ and $(S,\eta)$, that is
$$
\mbox{$d\<Z,
W^1\>_t=\rho_1\,dt$, $d\<Z,
W^2\>_t=\rho_2\,dt$,
$d\<Z,
W^3\>_t=\rho_3\,dt$.
}
$$
Correlations among the processes $V$, $r$ and $\eta$ can be surely inserted (see next Remark \ref{corr-VXbis}).

As done in Section \ref{sect-model}, we take into account the transformations \eqref{rX}-\eqref{Xphi} for the generalized OU processes: we set
\begin{equation}\label{rthetaX}
r_t   = \sigma_r X^r_t + \varphi^r_t\quad\mbox{and}\quad
\eta_t   = \sigma_\eta X^\eta_t + \varphi^\eta_t
\end{equation}
where
\begin{equation}\label{Xetaphi}
\begin{array}{ll}
\displaystyle
X^r_t  = - \kappa_r \int_0^tX^r_s\,ds + \,W^2_t,
&\displaystyle
\quad
\varphi^r_t=r_0e^{-\kappa_r t}+\kappa_r\int_0^t\theta_r(s)e^{-\kappa_r(t-s)}ds,\smallskip\\
X^\eta_t  = - \kappa_\eta \int_0^tX^\eta_s\,ds + \,W^3_t,
&\displaystyle
\quad
\varphi^\eta_t=\eta_0e^{-\kappa_\eta t}+\kappa_\eta\int_0^t\theta_\eta(s)e^{-\kappa_\eta(t-s)}ds.
\end{array}
\end{equation}
So, by considering the $\log$-price process, we reduce to the $4$-dimensional process $(Y,V,X^r,X^\eta)$ whose dynamics is given by
\begin{equation}\label{YVXrXeta-dyn}
\begin{array}{ll}
&dY_t=\mu_Y(V_t,X^r_t, X^\eta_t, t)dt+\sqrt{V_t}\, \big(\rho_1dW^1_t+\rho_2dW^2_t+\rho_3dW^3_t+\rho_4dW^4_t\big), \smallskip\\
&dV_t= \mu_V(V_t)dt+\sigma_V\sqrt{V_t}\,dW^1_t,\smallskip\\
&dX^r_t=\mu_{X^r}(X^r_t)dt+dW^2_t,\smallskip\\
&dX^\eta_t=\mu_{X^\eta}(X^\eta_t)dt+dW^3_t,\smallskip\\
&\mbox{with}\quad
Y_0=\ln S_0\in\R,\quad
V_0>0,\quad
X^r_0=0,\quad
X^\eta_0=0
\end{array}
\end{equation}
where
\begin{align*}
&\rho_4=\sqrt{1-\rho_1^2-\rho_2^2-\rho_3^2}, \quad \mbox{with}\quad \rho_1^2+\rho_2^2+\rho_3^2<1,\\
&\mu_Y(v,x_1,x_2,t)=\sigma_rx_1+\varphi^r_t-\sigma_\eta x_2-\varphi^\eta_t-\frac 12 \,v,\\
&\mu_V(v)=\kappa_V(\theta_V-v),\quad
\mu_{X^r}(x)= -\kappa_r x,\quad \mu_{X^\eta}(x)= -\kappa_\eta x.
\end{align*}

Starting from \eqref{YVXrXeta-dyn}, we set-up an approximating procedure similar to the one developed in Section \ref{sect-tree} and Section \ref{sect-pde}.
In the following, we briefly describe how to extend such algorithms to the Heston-Hull-White2d  model.

\subsection{Approximation of $(V,X^r,X^\eta)$}\label{sect-apprVXrXeta}

Concerning the triple $(V,X^r,X^\eta)$, we build an approximating tree on $\R^3$ as follows:
\begin{itemize}
\item
we apply the procedure in Section \ref{sect-treeX} to the process $X^r$;
\item
we apply the procedure in Section \ref{sect-treeX} to the process $X^\eta$;
\item
we apply the procedure in Section \ref{sect-treeV} to the process $V$.
\end{itemize}
We then get three approximating trees:
$$
\mbox{$\hat X^{r,h}$ for $X^r$,\quad
$\hat X^{\eta,h}$ for $X^\eta$,\quad
$\hat V^{h}$ for $V$.
}
$$
Then, we use the null correlation between any two of $V$, $X^r$ and $X^\eta$:
we concatenate the above trees  in order to get a $3$-dimensional approximating tree
$(\hat V^{h},\hat X^{r,h},\hat X^{\eta,h})$ for $(V,X^r,X^\eta)$ by introducing product-type jump probabilities. In other words, we generalize the probabilities in \eqref{treescheme} for all the
$2^3=8$ possible jumps.

\begin{remark}\label{corr-VXbis}
One might include correlations between any two of the Brownian motions driving the processes $V$, $X^r$ and $X^\eta$. As described in Remark \ref{corr-VX}, the jump probabilities are no more of a product-type but they solve a linear system of equations that must include the matching of the local cross-moments up to order one in $h$.
\end{remark}

\subsection{The scheme on the $Y$-component and the approximating $4$-dimensional process}

We repeat the reasonings in Section \ref{sect-apprYVX} in order to define an approximating time-continuous process $(\bar Y^h,\bar V^{h},\bar X^{r,h},$ $\bar X^{\eta,h})$ for $(Y,V,X^r,X^\eta)$ - roughly speaking, it suffices to replace the one-dimensional process $X$ in Section \ref{sect-apprYVX} with the $2$-dimensional process $(X^r,X^\eta)$. So, we start from
\begin{equation}\label{Y1bis}
dY_t=\frac{\rho_1}{\sigma_V}dV_t+\rho_2\sqrt{V_t}dX^r_t
+\rho_3\sqrt{V_t}dX^\eta_t
+\mu(V_t,X^r_t,X^\eta_t,t)dt+\rho_4\sqrt{V_t}\,dW^4_t
\end{equation}
with
\begin{equation}\label{mu-finbis}
\mu(v,x_1,x_2,t)
=\mu_Y(v,x_1,x_2,t)-\frac{\rho_1}{\sigma_V}\mu_V(v)-\rho_2\sqrt{v}\,\mu_{X^r}(x_1)
-\rho_3\sqrt{v}\,\mu_{X^\eta}(x_2).
\end{equation}
Then, we apply the finite-difference method in Section \ref{sect-FD} and we obtain a final difference scheme given by
$$
\Pi(v,x_1,x_2)=A^{-1}(v,x_1,x_2)
$$
where, $\mu(\cdot)$ being defined in \eqref{mu-finbis} and $A$ is given in \eqref{matrixA} with
\begin{equation}\label{alphabeta_implbis}
\alpha=\frac{h}{\Dy}\,\mu(v,x_1,x_2,nh)\quad\mbox{and}\quad\beta=\frac{h}{2\Dy^2}\,\rho_4^2v.
\end{equation}
Finally, we extend the approximation scheme \eqref{E4} to the case in which $X=(X^r,X^\eta)$ and the algorithm for the pricing of European or American options described in Section \ref{sect-alg}.
\begin{remark}\label{complexity}
Let us briefly discuss the complexity of our algorithms. At each time step $n=0,\ldots,N=T/h$ one has to find the solution of a PDE on a grid with $2M+1$ points for each fixed values of
\begin{itemize}
\item
case 1, Heston-Hull-White model: the pair $(v,x)\in\mathcal{V}^h_n\times \mathcal{X}^h_n$,
 \item
case 2, Heston-Hull-White2d model: the triple $(v,x_1,x_2)\in\mathcal{V}^h_n\times \mathcal{X}^h_n\times \mathcal{X}^h_n$.
\end{itemize}
The cardinality of all these possible values in case $i$ is at most $n\times n^i$, $i=1,2$.
For each case, the system of equations \eqref{lin_sys} with tridiagonal matrix \eqref{matrixA}, can be solved by an efficient form of Gaussian elimination requiring a linear cost of order $O(M)$. Therefore, the total cost of our approach is of order
$$
\sum_{n=1}^{N}n^{i+1}\times (2M+1)
=O(N^{i+2}\times M),\quad\mbox{case $i=1,2$}.
$$
We notice that the use of a full finite-difference scheme could be more expensive for practical computations.
Indeed, consider case 1 (Heston-Hull-White model). The solution of a $3$-dimensional problem by applying finite-differences in all three components leads to the inversion of a big band-matrix. To reduce the computational cost, the problem requires to apply appropriate techniques such as ADI (Alternating Direction Implicit) techniques, see \cite{hh} and references therein. Specifically, in \cite{hh} the authors propose an ADI approach to solve the Heston-Hull-White partial differential equation which needs a non-trivial implementation effort with a computational cost at least of order $O(M^3)$ per time step, so the total cost is of order $O\big(N\times M^3\big)$ . Furthermore, as the dimension of the problem  increases, it is not clear what happens if the problem is solved with a full finite-difference scheme. In case 2 (Heston-Hull-White2d model), one should solve a 4-dimensional problem, bringing to the inversion of a very big band-matrix. This would give a cost which is hard to be quantified, and possibly in such a case the costs of the two procedures are no longer comparable.
\end{remark}

\section{The hybrid Monte Carlo algorithm}\label{sect-MC}

The approximation we have set-up for the Heston-Hull-White processes can be used to construct a Monte Carlo algorithm. Let us see how one can simulate a single path by using the tree approximation  and the standard Euler scheme for the $Y$-component. We call it ``hybrid'' because two different noise sources are considered: we simulate a continuous process in space (the component $Y$) starting from a discrete process in space (the 3-dimensional tree for $(V,X^r,X^\eta)$).

Concerning the Heston-Hull-White dynamics in Section \ref{sect-model}, consider the triple $(Y,V,X)$ as in \eqref{YVX-dyn}. Let $(\hat V^h_n,\hat X^h_n)_{n=0,1,\ldots,N}$ denote the Markov chain that approximates the pair $(V,X)$. We construct a sequence $(\hat Y_n)_{n=0,1,\ldots,N}$ approximating $Y$ at times $n=0,1,\ldots,N$ by means of the Euler scheme defined in \eqref{barYh}: we set
$\hat Y^h_0=Y_0$ and for $t\in[nh,(n+1)h]$ with $n=0,1,\ldots,N-1$ then
\begin{equation}\label{barYh-bis}
\hat Y^h_{n+1}
=\hat Y^h_{n}+
\frac{\rho_1}{\sigma_V}(\hat V^h_{n+1}-\hat V^h_{n})
+\rho_2\sqrt{\hat V^h_{n}}(\hat X^h_{n+1}-\hat X^h_{n})+\mu(\hat V^h_{n},\hat X^h_{n},{nh} )h+\rho_3\sqrt{h\hat V^h_{n}}\,\Delta_{n+1},
\end{equation}
where $\mu$ is defined in \eqref{mu-fin} and $\Delta_1,\ldots,\Delta_N$ denote i.i.d. standard normal r.v.'s, independent of the noise driving the chain $(\hat V,\hat X)$. So, the simulation algorithm is very simple: at each time step $n\geq 1$, one let the pair $(V,X)$ evolve on the tree and simulate the process $Y$ at time $nh$ by using \eqref{barYh-bis}.

A similar algorithm can be considered to simulate the Heston-Hull-White2d dynamics in Section \ref{sect-model2d}, that can be seen as a function of the triple $(Y,V,X^r, X^\eta)$ in \eqref{YVXrXeta-dyn}. Here, we apply the Euler scheme to \eqref{Y1bis}. So, let
$(\hat V^h_n,\hat X^{r,h}_n,\hat X^{\eta,h}_n)_{n=0,1,\ldots,N}$ denote the Markov chain approximating $(V,X^r, X^\eta)$, as described in Section \ref{sect-apprVXrXeta}. Starting from \eqref{Y1bis}, we approximate the component $Y$ at times $nh$, $n=0,1,\ldots,N$, as follows: we set $\hat Y^h_0=Y_0$ and for $n=1,\ldots,N$, $n=0,1,\ldots,N-1$ then
\begin{equation}\label{barY1h-bis}
\begin{array}{l}
\displaystyle
\hat Y^h_{n+1}
=\hat Y^h_{n}+
\frac{\rho_1}{\sigma_V}(\hat V^h_{n+1}-\hat V^h_n)+\rho_2\sqrt{\hat V^h_n}(\hat X^{r,h}_{n+1}-\hat X^{r,h}_n)+\rho_3\sqrt{\hat V^h_n}(\hat X^{\eta,h}_{n+1}-\hat X^{\eta,h}_n)\smallskip\\
\displaystyle
\quad +\mu(\hat V^h_n,\hat X^{r,h}_n,\hat X^{\eta,h}_n,nh)h+\rho_4\sqrt{h\hat V^h_n}\,\Delta_{n+1}
\end{array}
\end{equation}
where $\mu$ is defined in \eqref{mu-finbis} and $\Delta_1,\ldots,\Delta_N$ denote i.i.d. standard normal r.v.'s, independent of the noise driving the chain $(\hat V^h,\hat X^{r,h},\hat X^{\eta,h})$. And again, the simulation algorithm is straightforward.

\section{Numerical results}\label{sect-numerics}
In this section we provide numerical results in order to asses the
efficiency and the robustness of our hybrid numerical approach.
%
We first consider test experiments for the Heston-Hull-White model for the computation of European, American and barrier options (Section \ref{sect-num-euam}) and, following Andersen \cite{an}, we study Vanilla options with large maturities when the Feller condition is not fulfilled (Section \ref{andersen}). Then we test European and American options in the Heston-Hull-White2d model (Section \ref{sect-num-euam2d}).

\subsection{European, American and barrier options in the Heston-Hull-White
  model}\label{sect-num-euam}
In the European and American option contracts we are dealing with, we
consider the following set of parameters:
\begin{itemize}
\item
initial share price $S_0=100$, strike price $K=100$,
maturity $T=1$, dividend rate $\eta=0.03$;
\item
initial interest rate
$r_0=0.04$, speed of mean-reversion   $\kappa_r=1$, interest rate volatility  $\sigma_r=0.2$, time-varying long-term mean $\theta_r(t)$ which fits the theoretical bond prices to the yield curve observed on the market - to this purpose,
we have chosen  the interest rate curve given by $P_r(0,T)=e^{-0.04T}$;
\item
initial volatility $V_0=0.1$, long-mean $\theta_V=0.1$, speed of
mean-reversion   $\kappa_V=2$, volatility of volatility $\sigma_V=0.3$;
\item
varying correlations: for the pairs
$(S,V)$, and $(S,r)$, we set $\rho_1=\rho_{SV}=-0.5$ and $\rho_2=\rho_{Sr}=-0.5,0,0.5$ respectively; no correlation is assumed to exist between $r$ and $V$.
\end{itemize}

\noindent
We notice that, under the above requests, the Feller condition holds. We postpone to next Section \ref{andersen} the analysis of cases in which the Feller condition is not fulfilled.

The numerical study of the hybrid tree/finite-difference method
\textbf{HTFD} is split in two cases:
\begin{itemize}
\item[-]
\textbf{HTFD1} refers to the (fixed) number of time steps $N_t=50$ and varying number of space steps $N_S=50, 100, 150, 200$;
\item[-]
\textbf{HTFD2} refers to $N_t=N_S=50, 100, 150, 200$.
\end{itemize}
Concerning the Monte Carlo method, we compare the results by using the hybrid simulation scheme in Section \ref{sect-MC}, that we call \textbf{HMC}. We also simulate paths by using the accurate third-order Alfonsi \cite{al} discretization scheme for the CIR stochastic volatility process and by using an exact scheme for the
interest rate. These simulating schemes are here called \textbf{AMC}.
In both Monte Carlo methods, we consider  varying number of time discretization steps
$N_t=50, 100, 150, 200$ and two cases for the number of Monte Carlo
iterations:
\begin{itemize}
\item[-]
\textbf{HMC1} and \textbf{AMC1} refer to 50 000 iterations,
\item[-]
\textbf{HMC2} and \textbf{AMC2} refer to 200 000 iterations.
\end{itemize}

In the European case, the benchmark value \textbf{B-AMC} is computed using the Alfonsi method with $300$ discretization time steps and the associated Monte Carlo estimator is computed with 1 million simulations.
In the American case, in absence of reliable numerical methods, the benchmark values \textbf{B-AMC-LS} are obtained by the Longstaff-Schwartz \cite{ls} Monte Carlo algorithm with $50$
exercise dates, combined with the Alfonsi method with $300$ discretization time steps and 1 million iterations.

Table \ref{tab1} reports both European call
option prices and implied volatilities results.
In Table \ref{tab2} we provide American call option prices. Table \ref{tab5} refers to the
computational time cost (in seconds) of the different algorithms in
the call European case.

The numerical results show that \textbf{HTFD} is accurate, reliable
and efficient for pricing European and American
options in the Heston-Hull-White model.
Moreover, our hybrid Monte Carlo algorithm \textbf{HMC} appears to be competitive with \textbf{AMC}, that is the one from the accurate simulations by Alfonsi \cite{al}:
the numerical results are similar in term of precision
and variance but \textbf{HMC} is definitely  better from the computational times point of view. Additionally, because of its simplicity,
\textbf{HMC} represents a real and interesting alternative to \textbf{AMC}.
As a further evidence of the accuracy of our methods,
in Figure \ref{Fig1} we study the shapes of implied volatility smiles across
moneyness $\frac{K}{S_0}$ using \textbf{HTFD1} with $N_t=50$ and
$N_S=200$ and \textbf{HMC1} with $N_t=50$, and we compare the graphs
with the results from the benchmark.

In order to study the convergence behavior of  our approach HTFD,  we consider the convergence ratio proposed in \cite{dfl}, defined as
\begin{equation}\label{ratio}
\mathrm{ratio}=\frac{P_{\frac{N}{2}}-P_{\frac{N}{4}}}{P_{N}-P_{\frac{N}{2}}},
\end{equation}
where $P_{N}$ denotes here the approximated price obtained with
$N=N_t$ number of  time steps. Recall that $P_{N}=O(N^{-\alpha})$ means that
$\mathrm{ratio}=2^{\alpha}$. For the sake of comparison with the
numerical convergence speed studied in \cite{bcz}, we report ratios
for American put options. We split the numerical study in two
different cases: when the Feller condition holds and when it does not,
the results being given in Table \ref{tab2ratio} and Table
\ref{tab4ratio} respectively  (details on the option parameters are
given in the table captions). Both tables give evidence of the
numerical convergence, but with some differences. In fact, under the
Feller condition (Table \ref{tab2ratio}), the numerical speed of
convergence is definitely linear (this is not really surprising
because tree methods are usually linear), whereas in the opposite case
(Table \ref{tab4ratio}) the behavior is approximately linear.

\begin{table}[ht]
\centering
\subtable[]{
\footnotesize
{\begin{tabular} {@{}cccccccccc@{}} \toprule & $N_S$ &HTFD1
    &HTFD2 &B-AMC &HMC1 &HMC2 &AMC1 &AMC2\\
\hline
$\rho_{Sr}=-0.5$& 50&11.202744 &11.202744  &11.34$\pm$0.04
&11.30$\pm$0.16 &11.32$\pm$0.08 &11.34$\pm$0.16&11.37$\pm$0.08\\
&100 &11.319814 &11.331040 & &11.41$\pm$0.16 &11.38$\pm$0.08 &11.31$\pm$0.16&11.36$\pm$0.08\\
&150 &11.340665 &11.349902 & &11.36$\pm$0.16 &11.36$\pm$0.08 &11.35$\pm$0.16&11.38$\pm$0.08
\\
&200 &11.346972 &11.355772 & &11.34$\pm$0.16 &11.37$\pm$0.08 &11.44$\pm$0.16&11.39$\pm$0.08
\\
 \hline
$\rho_{Sr}=0 $& 50&12.526779 &12.526779
&12.77$\pm$0.04&12.66$\pm$0.18 &12.69$\pm$0.09 &12.68$\pm$0.18&12.79$\pm$0.09
\\
&100 &12.720651 &12.705772 & &12.74$\pm$0.18 &12.79$\pm$0.09 &12.63$\pm$0.18&12.78$\pm$0.09
\\
&150 &12.754610 &12.749526 & &12.74$\pm$0.18 &12.79$\pm$0.09 &12.68$\pm$0.18&12.81$\pm$0.09
\\
&200 &12.760365 &12.766836  & &12.74$\pm$0.18 &12.80$\pm$0.09 &12.75$\pm$0.18&12.79$\pm$0.09
\\
\hline
$\rho_{Sr}=0.5 $& 50&13.853193 &13.853193  &14.04$\pm$0.04
&13.88$\pm$0.19 &13.92$\pm$0.10 &13.97$\pm$0.20&14.05$\pm$0.10
\\
&100 &14.011537 &14.013063 & &13.91$\pm$0.19 &14.01$\pm$0.10 &13.89$\pm$0.19&14.06$\pm$0.10
\\
&150 &14.031598 &14.038361 & &13.94$\pm$0.19 &14.07$\pm$0.10 &13.92$\pm$0.20&14.08$\pm$0.10
\\
&200 &14.038235 &14.045612 & &13.99$\pm$0.19 &14.07$\pm$0.10 &13.90$\pm$0.19&14.06$\pm$0.10
\\
\hline
\end{tabular}}
}
\quad\quad
\subtable[]{
\footnotesize
{\begin{tabular} {@{}cccccccccc@{}} \toprule & $N_S$ &HTFD1
    &HTFD2 &B-AMC  &HMC1 &HMC2 &AMC1 &AMC2\\
\hline
$\rho_{Sr}=-0.5$& 50&0.279002 &0.279002  &0.282602 &0.281649 &0.282117 &0.282602 &0.283389\\
&100 &0.282073 &0.282367  & &0.284443 &0.283681  &0.281815 &0.283127\\
&150 &0.282620 &0.282862  & &0.283034 &0.283085  &0.282865 &0.283652\\
&200 &0.282785 &0.283016  & &0.282478 &0.283408  &0.285226 &0.283914\\
\hline
$\rho_{Sr}=0$& 50&0.313772 &0.313772  &0.320169 &0.317398 &0.317958  &0.317802 &0.320695\\
&100 &0.318871 &0.318480  & &0.319306 &0.320650  &0.316487 &0.320432\\
&150 &0.319764 &0.319630  & &0.319063 &0.320716  &0.317802 &0.321221\\
&200 &0.319916 &0.320086  & &0.319288 &0.321009  &0.319643 &0.320695\\
\hline
$\rho_{Sr}=0.5$& 50&0.348697 &0.348697  &0.353623 &0.349329 &0.350359  &0.351777 &0.353887\\
&100 &0.352873 &0.352913  & &0.350234 &0.352954  &0.349667 &0.354151\\
&150 &0.353402 &0.353580  & &0.350960 &0.354324  &0.350458 &0.354679\\
&200 &0.353577 &0.353771  & &0.352184 &0.354545  &0.349931 &0.354151\\
\hline
\end{tabular}}
}
\caption{\em \small{Prices (a)  and Implied volatilities (b) of European call options. $S_0=100$,
    $K=100$, $T=1$, $r_0=0.04$, $\kappa_r=1$, $\sigma_r=0.2$, $\eta=0.03$, $V_0=0.1$,
    $\theta_V=0.1$, $\kappa_V=2$, $\sigma_V=0.3$,
    $\rho_{Sr}=-0.5,0,0.5$, $\rho_{SV}=-0.5$.}}
\label{tab1}
\end{table}

\begin{table}[ht]
\centering
\footnotesize
{\begin{tabular} {@{}ccccc@{}} \toprule  & $N_S$ &HTFD1
    &HTFD2 & B-AMC-LS \\
\hline
$\rho_{Sr}=-0.5$& 50&12.090433 &12.090433  &12.22$\pm$0.01\\
&100 &12.205014 &12.212884  & \\
&150 &12.224432 &12.231392  & \\
&200 &12.230288 &12.237054  & \\
\hline
$\rho_{Sr}=0 $& 50&12.912708 &12.912708  &13.16$\pm$0.02\\
&100 &13.119121 &13.101073  & \\
&150 &13.156492 &13.149182  & \\
&200 &13.162893 &13.168602  & \\
\hline
$\rho_{Sr}=0.5 $& 50&13.944266 &13.944266  &14.15$\pm$0.02\\
&100 &14.125059 &14.122918  & \\
&150 &14.146240 &14.152060  & \\
&200 &14.153288 &14.160288  & \\
\hline
\end{tabular}
}
\caption{\em \small{Prices of American call options. $S_0=100$,
    $K=100$, $T=1$, $r_0=0.04$, $\kappa_r=1$, $\sigma_r=0.2$, $\eta=0.03$, $V_0=0.1$,
    $\theta_V=0.1$, $\kappa_V=2$, $\sigma_V=0.3$,
    $\rho_{Sr}=-0.5,0,0.5$, $\rho_{SV}=-0.5$.}}
\label{tab2}
\end{table}


\begin{table} [ht] \centering
\footnotesize
{\begin{tabular} {@{}ccccccccc@{}} \toprule & $N_S$ &HTFD1 &HTDF2    & B-AMC &HMC1 &HMC2  &AMC1 &AMC\\
\hline
& 50  &0.41 &0.41 &223.67 &0.77 &3.05 &2.16 &7.48\\
& 100 &0.84 &11.33 & &1.59 &6.11 &4.00 &14.61\\
& 150 &1.37 &49.99 & &2.33 &9.13 &5.87 &21.64\\
& 200  &1.87 &213.06 & &3.11 &12.73 &7.61 &28.85\\
\end{tabular}}
\caption{\em \small{Computational times (in seconds) for European call
    options.}}
\label{tab5}
\end{table}

Furthermore, we study the behavior of \textbf{HTFD} in the case of exotic options, namely for continuously
monitored barrier options. We consider call up-and-out options, whose payoff is given by
$$
(S_T-K)_+\I_{\{S_t<H\,\forall\,t\leq T\}}.
$$
In our numerical experiments, the up barrier is set at $H = 130$ and we
choose different values for $S_0 = 80,100,120$.
Table \ref{tab2b} reports European call up-and-out option prices. In the
barrier option case, we compare with a benchmark value, called
\textbf{B-AMC}, computed by 2 millions iterations which use the
Alfonsi \textbf{AMC} method with $9600$ discretization time steps.
The numerical results confirm the reliability of \textbf{HTFD} for barrier options.

\begin{figure}[ht]
\begin{center}
\includegraphics[scale=0.4]{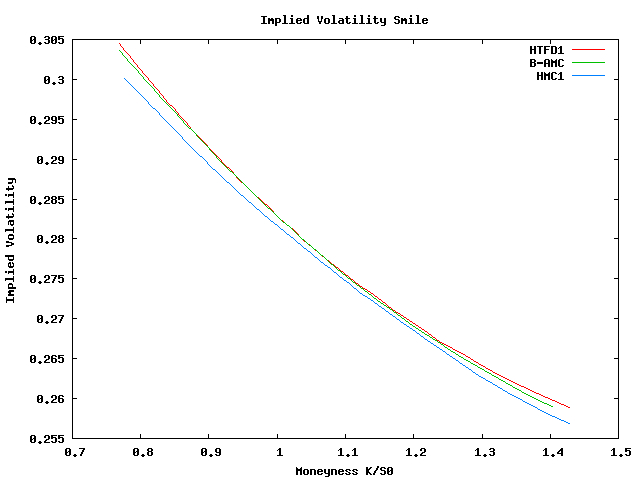}
\caption{Moneyness vs implied volatility for European call options. $T=1$, $r_0=0.04$, $\kappa_r=1$, $\sigma_r=0.2$, $\eta=0.03$, $V_0=0.1$,
    $\theta_V=0.1$, $\kappa_V=2$, $\sigma_V=0.3$,
    $\rho_{Sr}=-0.5$, $\rho_{SV}=-0.5$.}
\label{Fig1}
\end{center}
\end{figure}

\begin{table} [ht] \centering
 \footnotesize
 {\begin{tabular} {@{}ccccc@{}}\toprule K & $Nt$ & $N$ & Price & Ratio\\
  \hline
80 & 25 & 50 &  21.494606 & \\
   & 50 & 100 & 21.534555 & \\
   & 100 & 200 & 21.553473 & 2.111762 \\
   & 200 & 400 & 21.563911 & 1.812303 \\
   & 400 & 800 & 21.569080  & 2.019428 \\
\hline\hline
100 & 25 & 50 &  12.607035 & \\
   & 50 & 100 & 12.749006 & \\
   & 100 & 200 & 12.815657 & 2.130053 \\
   & 200 & 400 & 12.845050 & 2.267634 \\
   & 400 & 800 &  12.859561 & 2.025489 \\
\hline\hline
120 & 25 & 50 &  21.444819 & \\
  & 50 & 100 & 21.539534 &  \\
   & 100 & 200 & 21.572106 & 2.907912 \\
   & 200 & 400 & 21.586338 & 2.288708 \\
   & 400 & 800 &  21.592706 & 2.234825 \\
  \end{tabular}}
 \caption{\em \small{HTFD-ratio \eqref{ratio} for the price of
     American put options at final time $T=0.25$. $S_0=100$, $\eta=0.03$, $r_0=0.04$,$k_r=1$, $\sigma_r=0.2$, $V_0=0.1$, $k_V=2$, $\theta_V=0.1$, $\sigma_V=0.3$, $\rho_{SV}=-0.5$, $\rho_{Sr}=0.5$. }}
 \label{tab2ratio}
 \end{table}

 \begin{table} [ht] \centering
 \footnotesize
 {\begin{tabular} {@{}ccccc@{}}\toprule K & $Nt$ & $N$ & Price & Ratio\\
  \hline
80   & 25 & 50 &  21.635830 & \\
   & 50 & 100 & 21.669504 &  \\
   & 100 & 200 & 21.688879 & 1.738049 \\
   & 200 & 400 & 21.700965 & 1.603169 \\
   & 400 & 800 &  21.710373 &  1.284610 \\
\hline\hline
100   & 25 & 50 &  10.649104 & \\
   & 50 & 100 & 10.762867 &  \\
   & 100 & 200 & 10.812709 & 2.282480 \\
   & 200 & 400 & 10.835512 & 2.185787 \\
   & 400 & 800 & 10.848349  &  1.776369 \\
\hline\hline
120   & 25 & 50 &  20.755654 & \\
   & 50 & 100 & 20.873859 &  \\
   & 100 & 200 & 20.908825 & 3.380584 \\
   & 200 & 400 & 20.919694 & 3.216994 \\
   & 400 & 800 & 20.924295  & 2.362300 \\
  \end{tabular}}
 \caption{\em \small{HTFD-ratio \eqref{ratio} for the price of
     American put options at final time $T=0.25$. $S_0=100$, $\eta=0.03$, $r_0=0.04$,$k_r=1$, $\sigma_r=0.2$, $V_0=0.09$, $k_V=1$, $\theta_V=0.09$, $\sigma_V=1$, $\rho_{SV}=-0.3$, $\rho_{Sr}=0$. }}
 \label{tab4ratio}
 \end{table}

\begin{table} [ht] \centering
\footnotesize
{\begin{tabular} {@{}ccccc@{}} \toprule & $N_S$ &HTFD1  &HTFD2 & B-AMC
    \\
\hline
$S_0=80$& 50&1.211544 &1.211544  & \\
&100 &1.251453 &1.255849  &1.282211$\pm0.01$\\
&150 &1.264327 &1.270193  & \\
&200 &1.269703 &1.274332  & \\
\hline
$S_0=100 $& 50&1.819848 &1.819848  & \\
&100 &1.941320 &1.916440  &1.947565$\pm$0.01\\
&150 &1.964666 &1.930681  & \\
&200 &1.974201 &1.933482  & \\
\hline
$S_0=120 $& 50&0.697718 &0.697718  & \\
&100 &0.749116 &0.725243  &0.728431$\pm$0.01\\
&150 &0.762224 &0.726872  & \\
&200 &0.766022 &0.725139  & \\

\end{tabular}}
\caption{\em \small{Prices of European call up-and-out options. Up
    barrier is $H=130$. $K=100$, , $T=1$, $r_0=0.04$, $\kappa_r=1$, $\sigma_r=0.2$, $\eta=0.03$, $V_0=0.1$,
    $\theta_V=0.1$, $\kappa_V=2$, $\sigma_V=0.3$,
    $\rho_{Sr}=-0.5$, $\rho_{SV}=-0.5$.}}
\label{tab2b}
\end{table}

\clearpage
\subsection{European options with large maturity in the
  Heston-Hull-White model}\label{andersen}
In order to verify the robustness of the proposed algorithms we consider
experiments when the Feller condition is not fulfilled and with large maturities.
We test here the cases  I, II, III (reordered with respect to the maturity) proposed in Andersen \cite{an} in order to price
European call options. Moreover, we add the case 
IV with
maturity $T=25$.

We consider the following values for the parameters of the model and for the maturity date:
\begin{itemize}
\item[]
\bf{Case I:} $V_0=0.09$, $\theta_V=0.09$, $\kappa_V=1$,
  $\sigma_V=1$, $\rho_{SV}=-0.3$, $T=5$;
\item[]
\bf{Case II:} $V_0=0.04$, $\theta_V=0.04$, $\kappa_V=0.5$,
  $\sigma_V=1$, $\rho_{SV}=-0.9$, $T=10$.
\item[]
\bf{Case III:} $V_0=0.04$, $\theta_V=0.04$, $\kappa_V=0.3$,
  $\sigma_V=0.9$, $\rho_{SV}=-0.5$, $T=15$.
\item[]
\bf{Case IV:} $V_0=0.04$, $\theta_V=0.04$, $\kappa_V=0.3$,
  $\sigma_V=0.9$, $\rho_{SV}=-0.5$, $T=25$.
\end{itemize}
We take into account varying strikes $K=70, 100, 140$. No correlation is
assumed to exist between $S$ and $r$, that is $\rho_{Sr}=0$, so we can
compare the results with the semi closed-form analytic formula ({\bf
  SCF}) for European call options which is available in \cite{hh}. We
use in particular the implementation of the semi closed-form analytic
formula provided in QuantLib \cite{QL}.
Moreover in all cases the interest rate parameters, the initial share value and the dividend are the same of Section \ref{sect-num-euam}:
\begin{itemize}
\item $S_0=100$, $\eta=0.03$;
\item $r_0=0.04$, $\kappa_r=1$,  $\sigma_r=0.2$.
\end{itemize}

In Tables 
\ref{tab3-1}, \ref{tab3-2}, \ref{tab3-3}, \ref{tab3-4}  we
provide European call option prices and implied volatility
results. The numerical results suggest that large maturities bring to a slight loss of accuracy for both  \textbf{HTFD} and
\textbf{HMC}, even if each method
provides a satisfactory approximation of the true option prices. It is worth noticing that for long maturities $T=5,15,25$ we have developed experiments with the same number of steps both in time ($N_t$) and space ($N_S$) as for $T=1$. So, the numerical experiments are not slower, and it is clear that one could achieve a better accuracy for larger values of $N_t$.

\begin{table}[ht]
\centering
\subtable[]{
\footnotesize
{\begin{tabular} {@{}cccccccccc@{}} \toprule & $N_S$ &HTFD1
    &HTFD2 & SCF &HMC1 &HMC2 &AMC1 &AMC2\\
\hline
$K=70$& 50&37.054163 &37.054163  &37.491811 &37.36$\pm$0.47 &37.32$\pm$0.23 &37.38$\pm0.47$ &37.31$\pm$0.23\\
&100 &37.392491 &37.395372  & &37.30$\pm$0.45 &37.52$\pm$0.24 &37.61$\pm0.46$ &37.61$\pm$0.24\\
&150 &37.480467 &37.521733  & &37.40$\pm$0.46 &37.58$\pm$0.24 &37.55$\pm0.47$ &37.58$\pm$0.24\\
&200 &37.546885 &37.570675  & &37.42$\pm$0.46 &37.48$\pm$0.23 &37.49$\pm0.51$ &37.60$\pm$0.24\\
\hline
$K=100$& 50&23.997806 &23.997806  &24.706195 &24.64$\pm$0.43 &24.58$\pm$0.21 &24.61$\pm0.43$ &24.54$\pm$0.21\\
&100 &24.537750 &24.540987  & &24.49$\pm$0.41 &24.76$\pm$0.21  &24.79$\pm0.43$ &24.81$\pm$0.22\\
&150 &24.669356 &24.684708  & &24.60$\pm$0.41 &24.81$\pm$0.22  &24.71$\pm0.42$ &24.78$\pm$0.22\\
&200 &24.747161 &24.766840  & &24.67$\pm$0.42 &24.70$\pm$0.21  &24.73$\pm0.47$ &24.82$\pm$0.22\\
\hline
$K=140$ & 50&13.672435 &13.672435  &14.324566 &14.33$\pm$0.38 &14.24$\pm$0.18 &14.22$\pm0.37$ &14.17$\pm$0.18\\
&100 &14.248533 &14.205762  & &14.11$\pm$0.35 &14.40$\pm$0.198 &14.40$\pm0.37$ &14.40$\pm$0.19\\
&150 &14.373163 &14.318446  & &14.21$\pm$0.36 &14.43$\pm$0.198 &14.33$\pm0.38$ &14.40$\pm$0.19\\
&200 &14.444183 &14.404071  & &14.31$\pm$0.36 &14.32$\pm$0.188 &14.31$\pm0.42$ &14.42$\pm$0.20\\
\hline\
\end{tabular}
}
}
\quad\quad
\subtable[]{
\footnotesize
\begin{tabular} {@{}cccccccccc@{}} \toprule & $N_S$ &HTFD1
    &HTFD2 & SCF &HMC1 &HMC2 &AMC1 &AMC2\\
\hline
$K=70$& 50&0.313372 &0.313372  &0.322137 &0.319432 &0.318614 & 0.319863 & 0.318541\\
&100 &0.320152 &0.320209  & &0.318384 &0.322782 & 0.324521 & 0.324567 \\
&150 &0.321910 &0.322734  & &0.320315 &0.323861 & 0.323277 & 0.323956\\
&200 &0.323236 &0.323711  & &0.320737 &0.321815 & 0.322027 & 0.324318\\
\hline
$K=100$& 50&0.296912 &0.296912 &0.306954 &0.306002 &0.305124 & 0.30564 & 0.304539 \\
&100 &0.304563 &0.304608  & &0.303947 &0.307727 & 0.308148 & 0.308367\\
&150 &0.306431 &0.306649  & &0.305385 &0.308440 & 0.307022 & 0.307959 \\
&200 &0.307536 &0.307815  & &0.306431 &0.306889 & 0.307262 & 0.308556\\
\hline
$K=140 $& 50&0.291198 &0.291198  &0.299737 &0.299844 &0.298690 & 0.298395 & 0.240057\\
&100 &0.298743 &0.298183  & &0.296939 &0.300702 & 0.240301 & 0.239857\\
&150 &0.300373 &0.299657  & &0.298282 &0.301138 & 0.299848 & 0.300773\\
&200 &0.301301 &0.300777  & &0.299533 &0.299736 & 0.299505 & 0.301033\\
\hline
\end{tabular}}
\caption{\em \small{Prices (a)  and Implied volatilities (b) of European call options. $S_0=100$,
    $T=5$, $r_0=0.04$, $\kappa_r=1$, $\sigma_r=0.2$, $\eta=0.03$, $V_0=0.09$,
    $\theta_V=0.09$, $\kappa_V=1$, $\sigma_V=1$,
    $\rho_{Sr}=0$, $\rho_{SV}=-0.3$, $K=70, 100, 140$.}}
\label{tab3-1}
\end{table}

\begin{table}[ht]
\centering
\subtable[]{
\footnotesize
{\begin{tabular} {@{}cccccccccc@{}} \toprule & $N_S$ &HTFD1
    &HTFD2 & SCF &HMC1 &HMC2 &AMC1 &AMC2\\
\hline
$K=70$& 50&33.702753 &33.702753  &34.101622 &33.59$\pm$0.23 &33.76$\pm$0.11  &34.12$\pm0.23$ &34.09$\pm$0.11\\
&100 &33.773407 &34.120510  & &33.98$\pm$0.23 &34.10$\pm$0.11  &34.25$\pm0.23$ &34.10$\pm$0.11\\
&150 &33.776196 &33.818752  & &33.61$\pm$0.23 &33.76$\pm$0.11  &34.02$\pm0.23$ &34.11$\pm$0.11\\
&200 &33.778268 &33.944743  & &33.91$\pm$0.23 &33.91$\pm$0.11  &34.00$\pm0.23$ &34.10$\pm$0.11\\
\hline
$K=100$& 50&22.540546 &22.540546  &23.140518 &22.57$\pm$0.21 &22.65$\pm$0.10  &23.14$\pm0.21$ &23.09$\pm$0.11\\
&100 &22.761622 &23.076646  & &22.95$\pm$0.21 &23.06$\pm$0.10  &23.26$\pm0.21$ &23.12$\pm$0.11\\
&150 &22.795766 &22.857113  & &22.68$\pm$0.21 &22.81$\pm$0.10  &23.04$\pm0.21$ &23.09$\pm$0.11\\
&200 &22.806087 &22.978809  & &22.96$\pm$0.21 &22.95$\pm$0.10  &23.02$\pm0.21$ &23.13$\pm$0.11\\
\hline
$K=140$ & 50&13.335726 &13.335726  &13.755466 &13.18$\pm$0.17 &13.21$\pm$0.08  &13.72$\pm0.17$ &13.68$\pm$0.09\\
&100 &13.510432 &13.726749  & &13.53$\pm$0.17 &13.62$\pm$0.09  &13.86$\pm0.18$ &13.72$\pm$0.09\\
&150 &13.528322 &13.553294  & &13.34$\pm$0.17 &13.46$\pm$0.09  &13.66$\pm0.17$ &13.73$\pm$0.09\\
&200 &13.530288 &13.639595  & &13.60$\pm$0.17 &13.60$\pm$0.09  &13.64$\pm0.17$ &13.76$\pm$0.09\\
\hline
\end{tabular}}
}
\quad\quad
\subtable[]{
\footnotesize
{\begin{tabular} {@{}cccccccccc@{}} \toprule & $N_S$ &HTFD1
    &HTFD2 & SCF &HMC1 &HMC2 &AMC1 &AMC2\\
\hline
$K=70$& 50&0.227844 &0.227844  &0.234811 &0.225850 &0.228795 & 0.235187 & 0.234676\\
&100 &0.229082 &0.235140  & &0.232714 &0.234866 & 0.237332 & 0.234768\\
&150 &0.229131 &0.229876  & &0.226245 &0.228809 & 0.233392 & 0.235042\\
&200 &0.229167 &0.232077  & &0.231443 &0.231474 & 0.232965  & 0.234723\\
\hline
$K=100$& 50&0.215548 &0.215548  &0.222789 &0.215951 &0.216908 & 0.222801 & 0.222156\\
&100 &0.218214 &0.222017  & &0.220435 &0.221799 & 0.224184 & 0.222572\\
&150 &0.218625 &0.219366  & &0.217216 &0.218739 & 0.221545 & 0.222656 \\
&200 &0.218750 &0.220835  & &0.220583 &0.220512  & 0.221377 & 0.222760\\
\hline
$K=140$ & 50&0.210662 &0.210662  &0.215154 &0.209030 &0.209283 & 0.214777 & 0.214386\\
&100 &0.212532 &0.214847  & &0.212764 &0.213679 & 0.216253 & 0.214726\\
&150 &0.212723 &0.212991  & &0.210669 &0.212018 & 0.214162 & 0.214846\\
&200 &0.212744 &0.213914  & &0.213460 &0.213506 & 0.213908 & 0.215215\\
\hline
\end{tabular}}
}
\caption{\em \small{Prices (a)  and Implied volatilities (b) of European call options. $S_0=100$,
    $T=10$, $r_0=0.04$, $\kappa_r=1$, $\sigma_r=0.2$, $\eta=0.03$, $V_0=0.04$,
    $\theta_V=0.04$, $\kappa_V=0.5$, $\sigma_V=1$,
    $\rho_{Sr}=0$, $\rho_{SV}=-0.9$, $K=70, 100, 140$.}}
\label{tab3-2}
\end{table}

\begin{table}[ht]
\centering
\subtable[]{
\footnotesize
{\begin{tabular} {@{}cccccccccc@{}} \toprule & $N_S$ &HTFD1
    &HTFD2 & SCF &HMC1 &HMC2 &AMC1 &AMC2\\
\hline
$K=70$& 50&32.872766 &32.872766  &33.182814 &33.17$\pm$0.31 &33.26$\pm$0.16 &33.13$\pm$0.31 &33.18$\pm$0.16\\
&100 &33.041266 &33.161213  & &33.10$\pm$0.30 &33.29$\pm$0.15 &33.18$\pm$0.31 &33.19$\pm$0.15\\
&150 &33.098186 &33.159078  & &33.03$\pm$0.30 &33.12$\pm$0.16 &33.20$\pm$0.34 &33.25$\pm$0.16\\
&200 &33.150052 &33.235555  & &33.02$\pm$0.29 &33.11$\pm$0.15 &33.12$\pm$0.33 &33.35$\pm$0.15\\
\hline
$K=100$ & 50&24.738008 &24.738008  &25.183109 &25.00$\pm$0.30 &25.05$\pm$0.15 &25.10$\pm$0.30 &25.17$\pm$0.15\\
&100 &24.979024 &25.089961  & &24.96$\pm$0.29 &25.18$\pm$0.15 &25.20$\pm$0.30 &25.20$\pm$0.15\\
&150 &25.047214 &25.150207  & &24.99$\pm$0.28 &25.11$\pm$0.15 &25.17$\pm$0.33 &25.23$\pm$0.16\\
&200 &25.103492 &25.224136  & &24.97$\pm$0.28 &25.09$\pm$0.15 &25.07$\pm$0.31 &25.30$\pm$0.15\\
\hline
$K=140$ & 50&17.522401 &17.522401  &17.851374&17.49$\pm$0.27 &17.53$\pm$0.14  &17.76$\pm$0.27 &17.84$\pm$0.14\\
&100 &17.702990 &17.779408  & &17.51$\pm$0.26 &17.74$\pm$0.14 &17.85$\pm$0.28 &17.86$\pm$0.15\\
&150 &17.752550 &17.858103  & &17.59$\pm$0.26 &17.77$\pm$0.14  &17.82$\pm$0.31 &17.87$\pm$0.14\\
&200 &17.800293 &17.912261  & &17.59$\pm$0.25 &17.75$\pm$0.13 &17.73$\pm$0.29 &17.93$\pm$0.14\\
\hline
\end{tabular}}
}
\quad\quad
\subtable[]{
\footnotesize
{\begin{tabular} {@{}cccccccccc@{}} \toprule & $N_S$ &HTFD1
    &HTFD2 & SCF &HMC1 &HMC2 &AMC1 &AMC2\\
\hline
$K=70$& 50&0.231761 &0.231761 &0.237013  &0.236812 &0.238369 & 0.236053 & 0.236928\\
&100 &0.234617 &0.236648  & &0.235577 &0.238826 & 0.236974 & 0.237216 \\
&150 &0.235581 &0.236612  & &0.234478 &0.235946 & 0.237321 & 0.238081\\
&200 &0.236459 &0.237905  &  &0.234214 &0.235730  & 0.235877 & 0.239816\\
\hline
$K=100 $& 50&0.225390 &0.225390  &0.230837 &0.228633 &0.22926 & 0.229786 & 0.230708\\
&100 &0.228336 &0.229695  & &0.228045 &0.230802 & 0.231012 & 0.231068\\
&150 &0.229171 &0.230433  & &0.228424 &0.229935 & 0.230683 & 0.231375 \\
&200 &0.229861 &0.231340  &  &0.228236 &0.229682  &0.229474 & 0.232223\\
\hline
$K=140$& 50&0.223601 &0.223601  &0.227024  &0.223225 &0.223688  & 0.226081 & 0.226894\\
&100 &0.225479 &0.226275  &  &0.223424 &0.225813 & 0.227054 & 0.227136\\
&150 &0.225995 &0.227094  &   &0.224321 &0.226215 & 0.226728 & 0.227195\\
&200 &0.226492 &0.227659  &  &0.224316 &0.225961  & 0.225781  & 0.227881\\
\hline
\end{tabular}}
}
\caption{\em \small{Prices (a)  and Implied volatilities (b) of European call options. $S_0=100$,
    $T=15$, $r_0=0.04$, $\kappa_r=1$, $\sigma_r=0.2$, $\eta=0.03$, $V_0=0.04$,
    $\theta_V=0.04$, $\kappa_V=0.3$, $\sigma_V=0.9$,
    $\rho_{Sr}=0$, $\rho_{SV}=-0.5$, $K=70, 100, 140$.}}
\label{tab3-3}
\end{table}

\begin{table}[ht]
\centering
\subtable[]{
\footnotesize
{\begin{tabular} {@{}cccccccccc@{}} \toprule & $N_S$ &HTFD1
    &HTFD2 & SCF &HMC1 &HMC2 &AMC1 &AMC2\\
\hline
$K=70$& 50&28.772135 &28.772135  &28.969593 &29.01$\pm$0.29 &29.05$\pm$0.15 &28.92$\pm$0.29 &28.98$\pm$0.15\\
&100 &28.890859 &29.076376  & &29.06$\pm$0.34 &29.00$\pm$0.15 &28.99$\pm$0.29 &28.97$\pm$0.15\\
&150 &29.007171 &29.225059  & &29.07$\pm$0.29 &29.15$\pm$0.15 &28.90$\pm$0.31 &29.05$\pm$0.15\\
&200 &29.125812 &29.152251  & &29.05$\pm$0.31 &28.91$\pm$0.15 &28.95$\pm$0.29 &29.01$\pm$0.14\\
\hline
$K=100$& 50&23.947048 &23.947048  &24.255944 &24.09$\pm$0.28 &24.13$\pm$0.15 &24.20$\pm$0.29 &24.26$\pm$0.15\\
&100 &24.107443 &24.300298  & &24.22$\pm$0.33 &24.19$\pm$0.155 &24.27$\pm$0.28 &24.25$\pm$0.15\\
&150 &24.233382 &24.462163  & &24.26$\pm$0.28 &24.37$\pm$0.155 &24.17$\pm$0.31 &24.33$\pm$0.15\\
&200 &24.356051 &24.436578  & &24.32$\pm$0.31 &24.20$\pm$0.145 &24.22$\pm$0.28 &24.31$\pm$0.14\\
\hline
$K=140$ & 50&19.352114 &19.352114  &19.601699 &19.21$\pm$0.27 &19.24$\pm$0.14 &19.52$\pm$0.27 &19.59$\pm$0.14\\
&100 &19.459177 &19.637550  & &19.39$\pm$0.32 &19.39$\pm$0.14 &19.62$\pm$0.27 &19.59$\pm$0.14\\
&150 &19.567765 &19.778396  & &19.51$\pm$0.27 &19.62$\pm$0.14 &19.51$\pm$0.30 &19.66$\pm$0.14\\
&200 &19.692584 &19.798050  & &19.60$\pm$0.30 &19.53$\pm$0.14 &19.55$\pm$0.27 &19.65$\pm$0.13\\
\hline
\end{tabular}}
}
\quad\quad
\subtable[]{
\footnotesize
{\begin{tabular} {@{}cccccccccc@{}} \toprule & $N_S$ &HTFD1
    &HTFD2 & SCF &HMC1 &HMC2 &AMC1 &AMC2\\
\hline
$K=70$& 50&0.235972 &0.235972  &0.239830 &0.240620 &0.241466 & 0.238797  & 0.240057\\
&100 &0.238291 &0.241919  & &0.241551 &0.240401  & 0.240301 & 0.239857\\
&150 &0.240565 &0.244831  & &0.241698 &0.243433 & 0.239857 & 0.241365\\
&200 &0.242887 &0.243405  & &0.241403 &0.238618 & 0.239442 & 0.239442\\
\hline
$K=100$& 50&0.231127 &0.231127  &0.235633 &0.233270 &0.233862 & 0.234826  & 0.235633\\
&100 &0.233464 &0.236283  & &0.235133 &0.234655 & 0.235771 & 0.235536\\
&150 &0.235303 &0.238656  & &0.235636 &0.237270 & 0.234367 & 0.236650\\
&200 &0.237099 &0.238281  & &0.236537 &0.234765 & 0.235096 & 0.236425\\
\hline
$K=140$& 50&0.229635 &0.229635  &0.232641 &0.227954 &0.228325 & 0.231645 & 0.232511\\
&100 &0.230923 &0.233074  & &0.230084 &0.230128 & 0.233016  & 0.232455\\
&150 &0.2332231 &0.234777  & &0.231504 &0.232913 & 0.231583 & 0.233359\\
&200 &0.2333739 &0.235015  & &0.232634 &0.231717 & 0.232046 & 0.233281\\
\hline
\end{tabular}}
}
\caption{\em \small{Prices (a)  and Implied volatilities (b) of European call options. $S_0=100$,
    $T=25$, $r_0=0.04$, $\kappa_r=1$, $\sigma_r=0.2$, $\eta=0.03$, $V_0=0.04$,
    $\theta_V=0.04$, $\kappa_V=0.3$, $\sigma_V=0.9$,
    $\rho_{Sr}=0$, $\rho_{SV}=-0.5$, $K=70, 100, 140$.}}
\label{tab3-4}
\end{table}

\clearpage
\subsection{European and American options in the Heston-Hull-White2d model}\label{sect-num-euam2d}
In the European and American option contracts we are dealing with, we
consider the following set of parameters:
\begin{itemize}
\item $S_0=100$, $K=100$, $T=1$;
\item  $r_0=0.04$, $\eta_0=0.03$, $\kappa_r=\kappa_{\eta}=1$, $\sigma_r=\sigma_{\eta}=0.2$;
\item $V_0=0.1$, $\theta_V=0.1$, $\kappa_V=2$, $\sigma_V=0.3$;
\item $\rho_{Sr}=-0.5,0,0.5$, $\rho_{SV}=-0.5$,
  $\rho_{S\eta}=-0.5,0.5$,  $\rho_{Vr}=\rho_{V\eta}=\rho_{r\eta}=0$;
\item $P_r(0,T)=e^{-0.04T}$, $P_{\eta}(0,T)=e^{-0.03T}.$
\end{itemize}
As before, the
time-varying long-term means $\theta_r(t)$ and $\theta_\eta(t)$ fit the theoretical bond
prices $P_{r}(0,T)$ and $P_{\eta}(0,T)$ to the yield curve observed on the
market. We make this choice following the  multi-currency models with short-rate interest rates in \cite{go12}.
We consider here only the number of space steps $N_S=30, 50, 100$
because the cases $N_S=150, 200$ need a too high computational time.
Tables \ref{tab67-1}, \ref{tab67-2} and \ref{tab78} report European and American call
option prices and implied volatilities. As before, the benchmark value for European options is computed using the Alfonsi \textbf{B-AMC} method with $300$ discretization time steps and the associated Monte Carlo estimator is computed with 1 million iterations. Concerning the benchmark \textbf{B-AMC-LS} for American options, it is computed by means of  the
Longstaff-Schwartz \cite{ls} Monte Carlo algorithm with $50$
exercise dates, combined with the Alfonsi method with $300$ discretization time steps and 1 million iterations. Table \ref{tab9} refers to the computational time cost (in seconds) of
the different algorithms in the call European case.
In Figure \ref{Fig2}  we compare the shapes of implied volatility smiles across
moneyness $\frac{K}{S_0}$ using \textbf{HTFD1} with $N_t=30$ and
$N_S=100$ and \textbf{HMC1}  with $N_t=30$.
The numerical results confirm the good numerical behavior of \textbf{HTFD}  and
\textbf{HMC} in the Heston-Hull-White2d model as well.

\begin{table}[ht] \centering
\centering
\subtable[]{
\footnotesize
{\begin{tabular} {@{}cccccccccc@{}} \toprule $\begin{array}{l}\rho_{SV}=-0.5,\\\rho_{S\eta}=-0.5\end{array}$ & $N_S$ &HTFD1
    &HTFD2 & B-AMC &HMC1 &HMC2 &AMC1 &AMC2\\
\hline
$\rho_{Sr}=-0.5$& 30&13.470572 &13.470572  &13.79  $\pm$ 0.04
&13.82$\pm$0.20 &13.74$\pm$0.10 &13.83$\pm$0.20&13.79$\pm$0.10\\
&50 &13.688842 &13.671173 & &13.96$\pm$0.20 &13.81$\pm$0.10 &13.88$\pm$0.20&13.80$\pm$0.10
\\
&100 &13.790205 &13.781519 & &14.00$\pm$0.20 &13.80$\pm$0.10 &13.68$\pm$0.20&13.73$\pm$0.10
\\
\hline
$\rho_{Sr}=0 $& 30&14.736242 &14.736242  &15.04  $\pm$  0.05
&15.10$\pm$0.22 &14.99$\pm$0.11 &14.95$\pm$0.22&15.03$\pm$0.11
\\
&50 &14.958094 &14.946029 & &15.23$\pm$0.22 &15.04$\pm$0.11 &14.98$\pm$0.22&15.01$\pm$0.11
\\
&100 &15.019204 &15.032709 & &15.21$\pm$0.22 &15.04$\pm$0.11 &14.80$\pm$0.21&14.97$\pm$0.11
\\
\hline
$\rho_{Sr}=0.5 $& 30&15.805046 &15.805046  &16.19 $\pm$ 0.03
&16.13$\pm$0.23 &16.06$\pm$0.11 &16.04$\pm$0.23&16.17$\pm$0.12
\\
&50 &16.052315 &16.032043 & &16.33$\pm$0.23 &16.10$\pm$0.11 &16.09$\pm$0.23&16.13$\pm$0.12
\\
&100 &16.155354  &16.145308 & &16.24$\pm$0.23 &16.19$\pm$0.12 &15.93$\pm$0.23&16.12$\pm$0.12
\\
\hline
\end{tabular}}
}\quad\quad
\subtable[]{
\footnotesize
\begin{tabular} {@{}cccccccccc@{}} \toprule $\begin{array}{l}\rho_{SV}=-0.5,\\
\rho_{S\eta}=-0.5\end{array}$ & $N_S$ &HTFD1
    &HTFD2 & B-AMC &HMC1 &HMC2  &AMC1 &AMC2\\
\hline
$\rho_{Sr}=-0.5$& 30&0.338612 &0.338612  &0.347031 &0.347724 &0.345593
&0.348085 &0.347031\\
&50 &0.344364 &0.343898  & &0.351511 &0.347675 &0.349404 &0.347294\\
&100 &0.347036 &0.346807  & &0.352510 &0.347205 &0.344131 &0.345449\\
\hline
$\rho_{Sr}=0 $& 30&0.372004 &0.372004  &0.380033 &0.381610 &0.378689 &0.377653 &0.379769\\
&50 &0.377867 &0.377549  & &0.385153 &0.380032 &0.378447 &0.379240 \\
&100 &0.379483 &0.379840  & &0.384419 &0.380061 &0.373689 &0.378182\\
\hline
$\rho_{Sr}=0.5 $& 30&0.400281 &0.400281  &0.410485 &0.408889 &0.407043
&0.406508 &0.409954\\
&50 &0.406834 &0.406297  & &0.414273 &0.408039 &0.407833 &0.408894\\
&100 &0.409566 &0.409300  &  &0.411792 &0.410506 &0.403592 &0.408629\\
\hline
\end{tabular}}
\caption{\em \small{Prices (a) and Implied volatilities (b) of European call options. $S_0=100$,
    $K=100$, $T=1$, $r_0=0.04$, $\kappa_r=1$, $\sigma_r=0.2$, $\eta_0=0.03$, $\kappa_{\eta}=1$, $\sigma_{\eta}=0.2$, $V_0=0.1$,
    $\theta_V=0.1$, $\kappa_V=2$, $\sigma_V=0.3$,
    $\rho_{Sr}=-0.5,0,0.5$, $\rho_{SV}=-0.5$, $\rho_{S\eta}=-0.5$.}}
\label{tab67-1}
\end{table}

\begin{table}[ht] \centering
\centering
\subtable[]{
\footnotesize
{\begin{tabular} {@{}cccccccccc@{}} \toprule $\begin{array}{l}\rho_{SV}=-0.5,\\\rho_{S\eta}=0.5\end{array}$ & $N_S$ &HTFD1
    &HTFD2 & B-AMC &HMC1 &HMC2 &AMC1 &AMC2\\
\hline
$\rho_{Sr}=-0.5$& 30&9.418513 &9.418513  &9.61 $\pm$ 0.03
&9.57$\pm$0.13 &9.62$\pm$0.07 &9.64$\pm$0.13&9.66$\pm$0.07\\
&50 &9.552565 &9.532194 & &9.57$\pm$0.13 &9.61$\pm$0.07 &9.65$\pm$0.13&9.66$\pm$0.07
\\
&100 &9.633716 &9.607339 & &9.66$\pm$0.13 &9.62$\pm$0.07 &9.63$\pm$0.13&9.63$\pm$0.07
\\
\hline
$\rho_{Sr}=0 $& 30&10.916753 &10.916753  &11.18 $\pm$ 0.03
&11.15$\pm$0.15 &11.16$\pm$0.08 &11.07$\pm$0.15&11.22$\pm$0.08
\\
&50 &11.117050 &11.100343 & &11.18$\pm$0.15 &11.16$\pm$0.08 &11.14$\pm$0.15&11.22$\pm$0.08
\\
&100 &11.178119 &11.173631 & &11.16$\pm$0.15 &11.18$\pm$0.08 &11.08$\pm$0.15&11.20$\pm$0.08
 \\
\hline
$\rho_{Sr}=0.5 $& 30&12.203271 &12.203271  &12.55 $\pm$  0.04
&12.44$\pm$0.17 &12.43$\pm$0.09 &12.47$\pm$0.17&12.60$\pm$0.09
\\
&50 &12.443197 &12.411406 & &12.54$\pm$0.17 &12.44$\pm$0.09 &12.53$\pm$0.17&12.59$\pm$0.09
\\
&100 &12.552842 &12.522237 & &12.45$\pm$0.17 &12.55$\pm$0.09 &12.45$\pm$0.17&12.58$\pm$0.09
\\
\hline
\end{tabular}}
}\quad\quad
\subtable[]{
\footnotesize
\begin{tabular} {@{}cccccccccc@{}} \toprule $\begin{array}{l}\rho_{SV}=-0.5,\\
\rho_{S\eta}=0.5\end{array}$ & $N_S$ &HTFD1
    &HTFD2 & B-AMC &HMC1 &HMC2  &AMC1 &AMC2\\
\hline
$\rho_{Sr}=-0.5$& 30&0.232267 &0.232267  &0.237277 &0.236285 &0.237525
&0.238062 &0.238586\\
&50 &0.235774 &0.235241  &  &0.236157 &0.237263 &0.238324 &0.238586\\
&100 &0.237898 &0.237208  &  &0.238622 &0.237412 &0.237800 &0.237800\\
\hline
$\rho_{Sr}=0 $& 30&0.271502 &0.271502  &0.278405  &0.277704 &0.277855 &0.275520 &0.279454\\
&50 &0.276754 &0.276316  &  &0.278277 &0.277937 &0.277356 &0.279454\\
&100 &0.278356 &0.278238  &  &0.277841 &0.278435 &0.275782 &0.278930\\
\hline
$\rho_{Sr}=0.5 $& 30&0.305269 &0.305269  &0.314383  &0.311364
&0.311103 &0.312279 &0.315698\\
&50 &0.311575 &0.310739  &  &0.313992 &0.311570 &0.313857 &0.315435\\
&100 &0.314458 &0.313653  &  &0.311665 &0.314463 &0.311754 &0.315172\\
\hline
\end{tabular}}
\caption{\em \small{Prices (a) and Implied volatilities (b) of European call options. $S_0=100$,
    $K=100$, $T=1$, $r_0=0.04$, $\kappa_r=1$, $\sigma_r=0.2$, $\eta_0=0.03$, $\kappa_{\eta}=1$, $\sigma_{\eta}=0.2$, $V_0=0.1$,
    $\theta_V=0.1$, $\kappa_V=2$, $\sigma_V=0.3$,
    $\rho_{Sr}=-0.5,0,0.5$, $\rho_{SV}=-0.5$, $\rho_{S\eta}=0.5$.}}
\label{tab67-2}
\end{table}

\begin{table}[ht] \centering
\centering
\subtable[]{
\footnotesize
{\begin{tabular} {@{}ccccc@{}} \toprule $\begin{array}{l}\rho_{SV}=-0.5,\\ \rho_{S\eta}=-0.5\end{array}$ & $N_S$ &HTFD1
    &HTFD2 & B-AMC-LS \\
\hline
$\rho_{Sr}=-0.5$& 30&14.057963 &14.057963  &14.40 $\pm$ 0.02\\
&50 &14.290597 &14.263254  & \\
&100 &14.400377 &14.381552  & \\
\hline
$\rho_{Sr}=0 $& 30&14.989844 &14.989844  &15.32 $\pm$ 0.02\\
&50 &15.253011 &15.229151  & \\
&100 &15.320569 &15.331744  & \\
\hline
$\rho_{Sr}=0.5 $& 30&15.826696 &15.826696  &16.28 $\pm$ 0.02\\
&50 &16.146080 &16.111559  & \\
&100 &16.270439 &16.248656  & \\
\hline
\end{tabular}}
}\quad\quad
\subtable[]{
\footnotesize
{\begin{tabular} {@{}ccccc@{}} \toprule $\begin{array}{l}\rho_{SV}=-0.5,\\\rho_{S\eta}=0.5\end{array}$ & $N_S$ &HTFD1
    &HTFD2 & B-AMC-LS \\
\hline
$\rho_{Sr}=-0.5$& 30&11.598655 &11.598655  &11.72 $\pm$ 0.02\\
&50 &11.707669 &11.681873  & \\
&100 &11.775632 &11.743388  & \\
\hline
$\rho_{Sr}=0 $& 30&12.400256 &12.400256  &12.60 $\pm$  0.02\\
&50 &12.579124 &12.561214  & \\
&100 &12.634969 &12.629401  & \\
\hline
$\rho_{Sr}=0.5 $& 30&13.137621 &13.137621  &13.47 $\pm$   0.02\\
&50 &13.380571 &13.341882  & \\
&100 &13.497053 &13.459978  & \\
\hline
\end{tabular}}
}
\caption{\em \small{Prices of American call options. $S_0=100$,
    $K=100$, $T=1$, $r_0=0.04$, $\kappa_r=1$, $\sigma_r=0.2$, $\eta_0=0.03$, $\kappa_{\eta}=1$, $\sigma_{\eta}=0.2$, $V_0=0.1$,
    $\theta_V=0.1$, $\kappa_V=2$, $\sigma_V=0.3$,
    $\rho_{Sr}=-0.5,0,0.5$, $\rho_{SV}=-0.5$, $\rho_{S\eta}=-0.5,
    0.5$.}}
\label{tab78}
\end{table}

\begin{table} [ht] \centering
\footnotesize
{\begin{tabular} {@{}ccccccccc@{}} \toprule & $N_S$ &HTFD1 &HTDF2  & B-AMC &HMC1 &HMC2  &AMC1 &AMC2\\
\hline
& 30 &2.22 &2.22 &284.84 &0.60 &2.61 &1.79 &6.03\\
& 50 &4.15 &24.56 & &1.14 &4.19 &2.73 &9.58\\
& 100  &7.95 &998.1 & &2.02 &8.06 &5.05 &18.70\\
\end{tabular}}
\caption{\em \small{Computational times (in seconds) for European call
    options.}}
\label{tab9}
\end{table}


\clearpage

\begin{figure}[ht]
\begin{center}
\includegraphics[scale=0.4]{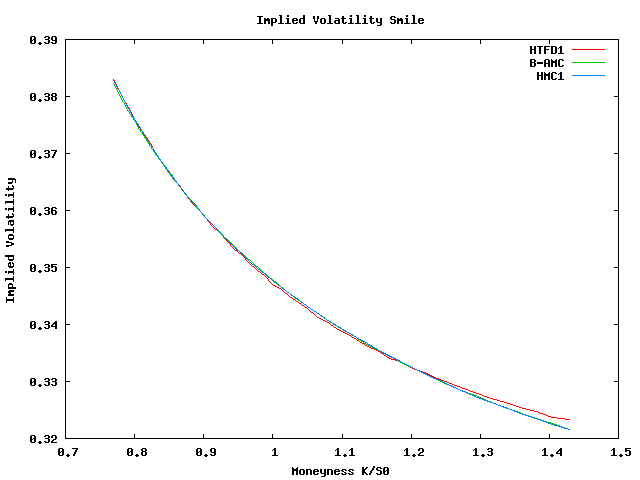}
\caption{Moneyness vs implied volatility for European call options. $T=1$, $r_0=0.04$, $\kappa_r=1$, $\sigma_r=0.2$, $\eta_0=0.03$, $\kappa_{\eta}=1$, $\sigma_{\eta}=0.2$, $V_0=0.1$,
    $\theta_V=0.1$, $\kappa_V=2$, $\sigma_V=0.3$,
    $\rho_{Sr}=-0.5$, $\rho_{SV}=-0.5$, $\rho_{S\eta}=-0.5$.}
\label{Fig2}
\end{center}
\end{figure}

\section{Conclusions}
We have introduced a new hybrid tree/finite-difference method and a new Monte Carlo method for numerically pricing options in
a stochastic volatility framework with stochastic interest rates.
The numerical comparisons show that our methods provide a good
approximation of the option prices with efficient time computations.

\medskip

\noindent
\textbf{Acknowledgements.}
The authors wish to thank Andrea Molent for useful remarks and for having implemented the
Alfonsi Monte Carlo scheme and the Longstaff-Schwarz algorithm.

\end{document}